\documentclass[fleqn,usenatbib]{mnras}

\usepackage{newtxtext,newtxmath,lscape}

\usepackage[T1]{fontenc}
\usepackage{ae,aecompl}

\usepackage{graphicx}	
\usepackage{amsmath}	
\usepackage{amssymb}	
\usepackage{multicol}

\title[To $\beta$ or not to $\beta$?]{To $\beta$ or not to $\beta$: can higher-order Jeans analysis break the mass--anisotropy degeneracy in simulated dwarfs?}
\author[A. Genina et al.]{A. Genina$^{1}$\thanks{E-mail: anna.genina@durham.ac.uk (AG)}, J. I. Read$^{2}$, C. S. Frenk$^{1}$, S. Cole$^{1}$,
A. Ben\'{i}tez-Llambay$^{1}$, \newauthor
A. D. Ludlow$^{3}$, J. F. Navarro$^{4\thanks{Senior CIfAR Fellow (JFN)}}$, K. A. Oman$^{1}$, A. Robertson$^{1}$ \\
$^{1}$ Institute for Computational Cosmology, Department of Physics, Durham University, South Road, Durham DH1 3LE, UK \\
$^{2}$ Department of Physics, University of Surrey, Guildford, GU2 7XH, UK \\
$^{3}$ International Centre for Radio Astronomy Research, University of Western Australia, 35 Stirling Highway,\\
\ \ Crawley,Western Australia, 6009, Australia \\
$^{4}$ Department of Physics \& Astronomy, University of Victoria, Victoria, BC, V8P 5C2, Canada }

\date{Accepted XXX. Received YYY; in original form ZZZ}

\pubyear{2020}

\begin{document}
\label{firstpage}
\pagerange{\pageref{firstpage}--\pageref{lastpage}}
\maketitle

\begin{abstract}
We test a non-parametric higher-order Jeans analysis method, {\sc GravSphere}, on 32 simulated dwarf galaxies comparable to classical Local Group dwarfs like Fornax. The galaxies are selected from the APOSTLE suite of cosmological hydrodynamics simulations with Cold Dark Matter (CDM) and Self-Interacting Dark Matter (SIDM) models, allowing us to investigate cusps and cores in density distributions. We find that, for CDM dwarfs, the recovered enclosed mass profiles have a bias of no more than 10~per~cent, with a 50~per~cent scatter in the inner regions and a 20~per~cent scatter near the half-light radius, consistent with standard mass estimators. The density profiles are also recovered with a bias of no more than 10~per~cent and a scatter of 30~per~cent in the inner regions. For SIDM dwarfs, the mass and density profiles are recovered within our 95~per~cent confidence intervals, but are biased towards cuspy dark matter distributions. This is mainly due to a lack of sufficient constraints from the data. We explore the sources of scatter in the accuracy of the recovered profiles and suggest a $\chi^2$ statistic to separate successful models from biased ones. Finally, we show that the uncertainties on the mass profiles obtained with {\sc GravSphere} are smaller than those for comparable Jeans methods, and that they can be further improved if stronger priors, motivated by cosmological simulations, are placed on the velocity anisotropy. We conclude that {\sc GravSphere} is a promising Jeans-based approach for modelling dark matter distributions in dwarf galaxies.
\end{abstract}




\section{Introduction}
\label{intro}
Dark matter makes up $\sim$85 per cent of the matter density of the Universe (e.g. \citealt{planck}); yet its identity remains unknown. Dwarf satellite galaxies of the Milky Way are expected to be excellent sites for testing the properties of dark matter \citep{battaglia_review}. These objects have velocity dispersions indicative of a high dark matter content. If dark matter is a self-annihilating particle, the products of its annihilation may be detected with space and ground-based instruments \citep{lake_dm}, such as Fermi-LAT \citep{fermi} and the upcoming Cherenkov Telescope Array \citep{cta}. The detected spectra of these annihilation products would reveal the particle physics properties of dark matter; however, these analyses require the underlying dark matter distribution to be well constrained. Some of the classical dwarf spheroidals, like Sculptor and Fornax, host $\sim$10$^6$--10$^8$M$_{\odot}$ in stars \citep{mcconnachieCensus} and are sufficiently close to our Galaxy that large samples of high-quality kinematic and photometric stellar data can be obtained, so inferences may be made of the underlying gravitational potential (e.g. \citealt{battagliasculptor,strigarinature,walkerest, charbonnier,walkerPenarrubia,bonnivard,heating,StrigariFrenkWhite}).

The exact dark matter density distribution in the central regions of dwarf galaxies is controversial (e.g. \citealt{corecuspreview}). Galaxy rotation curves measured in some dwarf irregulars \citep{Moorecore,ohcore,zhucores,adamsrotcurv,oh2015,read_curves,kyle}, analyses based on multiple tracer populations in dwarf spheroidals \citep{walkerPenarrubia,amoriscoevans} and globular cluster survivability within dwarfs \citep{goerdt,cole_glob_clust,contenta,orkney} have been used to argue in favour of dark matter `cores', where the density remains constant in the central regions, with $\rho \propto r^0$ \citep{floresprimack, Moorecore}. On the other hand, dark matter-only $N$-body simulations in $\Lambda$CDM cosmologies have found `cusps' in inner haloes, with density profiles scaling approximately as $\rho\propto r^{-1}$\citep{nfw96,nfw}. This became known as the `core-cusp problem'. The problem has motivated an introduction of alternative dark matter models, such as self-interacting dark matter, where cores are created through dark matter self-scattering on a scale related to the interaction cross-section \citep{selfintdm,sidmcores}. 

In recent years, the use of hydrodynamics to model baryonic processes has become more common in simulations. Outflows associated with supernova feedback have been shown to cause fluctuations in the gravitational potential, which can alter the inner structure of haloes. This may occur in a single violent burst \citep{navarroeke}, through repeated dark matter `heating' over time \citep{readgilmore,pontzen,firecores2,alltheway} or both \citep{alejandrocores}. Cores in these models typically form on the scale of the half-light radius of dwarf galaxies. 

The formation of cores in dwarfs that have undergone extended periods of star formation is a testable hypothesis. This idea was explored in \citet{heating}. These authors find that dwarf spheroidals which have continued to form stars until recent times, like Fornax and the Local Group dwarf irregulars, have lower densities at 150~pc, $\rho_{150}$, than those predicted for isolated dwarfs using the halo mass-concentration relation in $\Lambda$CDM from \citet{mass-conc}. These lower densities could be explained by core formation through dark matter `heating'. Dwarfs that have ceased star formation a long time ago have higher values of $\rho_{150}$, consistent with a cusp. 

The dark matter density distribution in dwarf galaxies can be constrained through Jeans analysis applied to line-of-sight stellar velocities and projected positions. This relies on the spherical Jeans equation:
\begin{equation}
\label{jeans}
\frac{1}{\nu}\frac{d}{d r}(\nu\sigma^2_r) + 2\frac{\beta \sigma^2_r}{r} = -\frac{GM(<r)}{r^2}, 
\end{equation}
where $\nu$ is the tracer number density distribution, $\sigma_r$ is their radial velocity dispersion, $\beta$ their velocity anisotropy, $M(<r)$ is the enclosed mass and $G$ is Newton's gravitational constant. The velocity anisotropy, $\beta$, is defined as $\beta =1 - \frac{\sigma^2_t}{2\sigma^2_r}$, where $\sigma_t$ is the tangential velocity dispersion. The product $\nu(r) \sigma^2_r(r)$ is typically obtained through deprojection of $\Sigma(R) \sigma^2_{P} (R)$, where $\sigma_{P} (R)$ is the line-of-sight velocity dispersion and $\Sigma(R)$ is the projected tracer number density at a distance, $R$, both of which are observable quantities. If models are assumed for $\beta(r)$ and $M(<r)$, the equation can be solved for $\sigma(R)$ via sampling methods such as Markov chain Monte Carlo (MCMC). This analysis assumes a non-rotating spherical system in a steady pseudo-equilibrium state. These assumptions are known to be violated by Local Group dwarfs, which exhibit ellipticity \citep{mcconnachieCensus}, signs of rotation \citep{battagliasculptor,delpino}, and are susceptible to tidal effects from their hosts \citep{readtides, penarrubiatides,disequilibrium}.

Typically, only the line-of-sight motions of the stellar tracers are known. This means that $\beta$ is poorly constrained, such that Jeans analysis suffers from the $M$--$\beta$ degeneracy. This degeneracy results in a wide range of models that satisfy a set of observational constraints, such that cored and cuspy dark matter profiles both provide acceptable fits to line-of-sight data \citep{sfw2010}. 

The breaking of  $M$ -- $\beta$ has been widely explored in the literature. Several works have focused on the use of multiple tracer populations in dwarf spheroidals and chemo-dynamical models \citep{walkerPenarrubia,amoriscoevans,agnelloevans,Zhu_2016}, as well as proper motions \citep{strigariproper,StrigariFrenkWhite}. A number of works have used Schwarzschild orbit superposition methods \citep{schwarz, Jardel_2012, Breddels_2013,kowal2017,kowal2018,schwarz_fornax}, which are able to take into account the asphericity of stellar systems and have a benefit of making no assumptions about the velocity anisotropy. These methods, however, typically require significant computing time.

Other works, based on the Jeans equation, have focused on exploiting the higher-order velocity moments. Specifically, it has been shown that line-of-sight velocity distributions are non-Gaussian in the absence of isotropy \citep{merritt1987}. This warrants the use of the 4$^{\rm th}$ moment of the velocity distribution to place a constraint on the anisotropy parameter \citep{mamposst}. The use of the fourth velocity moments has been explored for the case of constant anisotropy by \citet{lokas2002,lokasmamoncoma,lokasmamonprada,lokas2009} and was generalized for radially varying anisotropies by \citet{richardsonfairbairn2013}. More recently, \citet{geraint-fornax} have presented a non-parametric method of reconstruction of the line-of-sight velocity dispersion profiles. An extension of the method employs a machine learning approach for data reconstruction that proves useful in the absence of large samples of kinematic data \citep{geraint}. 

Another method for breaking this degeneracy is through the fourth order projected virial theorem, giving rise to two equalities \citep{merrifieldkent}:
\begin{equation}
\label{vsp1}
    \mathrm{VSP1}=\frac{2}{5}\int_{0}^{\infty}GM\nu(5-2\beta)\sigma^2_rrdr = \int_{0}^{\infty}\Sigma \langle \sigma^4_{P} \rangle RdR    
\end{equation}
and
\begin{equation}
\label{vsp2}
    \mathrm{VSP2}=\frac{4}{35}\int_{0}^{\infty}GM\nu(7-6\beta)\sigma^2_rr^3dr = \int_{0}^{\infty}\Sigma \langle \sigma^4_{P} \rangle R^3dR  ,
\end{equation}
where $\mathrm{VSP1}$ and $\mathrm{VSP2}$ are referred to as the virial shape parameters (i.e. VSPs). Here $\langle \sigma^4_{P} \rangle$ is the fourth moment of the line-of-sight velocities. The right-hand sides of Equations~\ref{vsp1} and~\ref{vsp2} contain quantities that can be directly inferred from data and the left-hand sides contain the same parameters as Equation~\ref{jeans}. It is thus possible to place two additional constraints on the velocity anisotropy $\beta$ (see e.g. \citealt{richardson}, where VSPs were used to show that a dark matter cusp is favoured in the Sculptor dwarf galaxy). In practice, however, the finite quality of data may result in only a partial breaking of the $M$ -- $\beta$ degeneracy.

\citet{gravsphere} introduced the non-parametric Jeans method, {\sc GravSphere} (used in \citealt{heating}), which employs the additional constraints from the VSPs in their MCMC analysis. {\sc GravSphere} operates under the standard assumptions of the spherical Jeans equation (spherical symmetry, equilibrium and no rotation). The method had been shown to recover successfully the dark matter density distributions in mock observations of idealized spherical, triaxial and tidally stripped simulated dwarfs from the {\sc GaiaChallenge} set\footnote{\url{http://astrowiki.ph.surrey.ac.uk/dokuwiki/}}. The cases for which the method works less well, such as aspherical systems, are evident through poor quality fits to the line-of-sight velocity dispersion. The method has been shown to recover accurately the densities at 150\,pc from the centre  -- a key region where core formation is expected to reduce dark matter densities, compared to $\Lambda$CDM predictions.

In this work, we test a new open source {\sc Python} implementation of the {\sc GravSphere} method, {\sc pyGravSphere}, on a sample of simulated dwarf galaxies from cosmological hydrodynamics simulations of Local Group-like environments\footnote{We ran also these same tests in the original {\sc GravSphere} code, yielding indistinguishable results from those presented here.}. The aim of our work is to establish whether Jeans analysis, under the assumption of spherical symmetry, is a suitable method for constraining the mass profiles of dark matter haloes in a fully cosmological setting and how much information is typically gained through the inclusion of VSPs. We explore the biases associated with mass and dark matter density profile recovery for individual dwarfs as well as the sample as a whole. We examine in detail the cases where {\sc GravSphere} fails and identify the reasons for this as well as potential warning signs.

 In Section~\ref{simulations}, we describe the suite of simulations used, as well as the setup of {\sc GravSphere}. Our analysis of {\sc GravSphere}'s performance on each galaxy and the comparison to more conventional Jeans methods can be found in Section~\ref{comparison}. We further discuss the various sources of bias in the performance of this method and suggest a $\chi^2$ statistic to weed out particularly biased models. We summarise our findings and conclude in Section~\ref{conclusions}.

\section{Simulations}
\label{simulations}

\subsection{APOSTLE simulations}

In this work, we select analogues of classical dwarf spheroidals from the APOSTLE suite of cosmological simulations. APOSTLE volumes are zoom-in simulations of Milky Way and Andromeda analogue pairs, selected from a dark matter-only volume. The pairs have been chosen to satisfy constraints for the Local Group, such as the total mass, separation and relative velocities. The details of halo selection may be found in \citet{fattahi}. Each Milky Way or Andromeda analogue hosts a population of dwarf galaxies. The volumes additionally include isolated dwarfs.

APOSTLE was run with the {\sc{eagle}} model of galaxy formation \citep{eagle, eaglecrain}, which is based on the smoothed particle hydrodynamics (SPH) $N$-body code {\sc gadget-3}, an improved version of the {\sc gadget-2} code \citep{gadget}. The original APOSTLE suite consists of five high-resolution cosmological volumes, with dark matter mass resolution of m$_{\rm{DM}} = 2.5-5\times10^4$M$_{\odot}$ and spatial resolution $\epsilon=134$~pc. An extra cosmological volume was also run assuming an SIDM interaction cross-section of $\sigma/m=10$~cm$^2$g$^{-1}$ \citep{santos,lovell}. We note that this is an extreme value of the cross-section, which was chosen to explore the formation of the largest cores in SIDM. The SIDM implementation within EAGLE was introduced in \citet{2018MNRAS.476L..20R}, based on the SIDM simulation method described in \citet{2017MNRAS.465..569R}. 

 In order to increase our sample of dwarfs with a dark matter core, we additionally used an SIDM version of the cosmological volume presented in \citet{alejandrocores}\footnote{The SIDM version of this volume has not yet been published. The simulation data were obtained through private communications.}, with an interaction cross-section of $\sigma / m = 10$~cm$^2$g$^{-1}$ and galaxy formation prescriptions following \citet{eagle} and \citet{eaglecrain}. This simulation was run with the same cosmological parameters as the APOSTLE simulations, but does not feature a Local Group-like setting. The dark matter particle mass resolution is m$_{\rm{DM}} = 4\times10^5$M$_{\odot}$ and the softening is $\epsilon= 234$~pc.

Each stellar particle in our simulations represents a stellar population assumed to follow a \citet{chabrier} initial mass function. Gas particles in APOSTLE have initial masses in the range $5-10\times10^3$M$_{\odot}$. APOSTLE resolves Sculptor-mass dwarf galaxies with $\sim$10$^2$-10$^3$ stellar particles and Fornax-mass dwarf galaxies with $\sim$10$^3$-10$^4$ particles. Further details of the APOSTLE simulations may be found in \citet{sawalapuzzles}, \citet{fattahi} and \citet{campbell}. The SIDM run of the volume presented in \citet{alejandrocores} has an initial gas particle mass m$_{\rm{gas}} = 6.6\times10^4$M$_{\odot}$. Fornax-mass dwarfs are resolved with $\sim10^2 - 10^3$ stellar particles, sufficient for the purposes of this work.

\subsection{Numerical considerations}

In order to establish whether {\sc GravSphere} reproduces the mass profiles of the simulated dwarfs, we must first define what the `true' mass profile is within the simulations. 

The mass profiles of dark matter haloes identified in pure $N$-body simulations are affected by collisional relaxation. The enclosed mass profiles of haloes are suppressed (relative to a higher resolution simulation) below a radius where the 2-body relaxation time, t$_{\rm relax}$, is comparable to the age of the Universe, $t_0$ (see e.g. \citealt{power,ludlowa}). For the typical number of dark matter particles in systems that are considered in this work, the \citet{power} radius, where t$_{\rm relax}\sim0.6t_{0}$, corresponds to $\sim$60 per cent of the half-light radius (about 0.7~kpc for high-resolution APOSTLE CDM and SIDM dwarfs and $\sim$~2~kpc for the lower-resolution SIDM volume from \citealt{alejandrocores}). Moreover, the APOSTLE simulations model dark matter and stars using particles of unequal mass, making them subject to energy equipartition, which artificially inflates galaxy sizes \citep{ludlowb, ludlowc}. These effects are most problematic for systems with stellar half-mass radii smaller than $\sim0.055$ of the mean interparticle separation (dark matter, stars and gas), corresponding to $\sim0.5$~kpc for the APOSTLE simulations and $\sim1.2$~kpc for the SIDM volume of \citet{alejandrocores}. We note, however, that the relaxation times at these radii are still considerably longer than the dynamical times of the stars. Therefore, our simulated dwarfs may be considered to be in a steady state locally. The ability of {\sc GravSphere} to recover masses in the innermost regions is of interest in this work; therefore we will present results below the convergence radius derived by \citet{power} and \citet{ludlowa}, although we will interpret these with caution\footnote{The \citet{power} criterion was derived for dark matter-only simulations and it is unclear how applicable this criterion is in the presence of a baryonic component.}.

The use of the gravitational softening in $N$-body simulations sets a limit on the central density, such that the innermost regions of simulated haloes exhibit a small artificial core on the scale of the gravitational softening. In comparing {\sc GravSphere} to $N$-body simulations, we thus restrict ourselves to radii greater than $2.8\epsilon$, where $\epsilon$ is the Plummer-equivalent gravitational softening. For the APOSTLE high-resolution simulations $\epsilon=134$~pc and $\epsilon=234$~pc for the SIDM version of the volume introduced in \citet{alejandrocores}\footnote{Although 2.8$\epsilon$ for this volume is near 0.5~kpc, the core sizes are typically much larger than that, such that we are still able to probe the interesting regions in these dwarfs.}. At the radius of $2.8\epsilon$ the forces become exactly Newtonian.

\begin{figure*}

		\includegraphics[width=2\columnwidth]{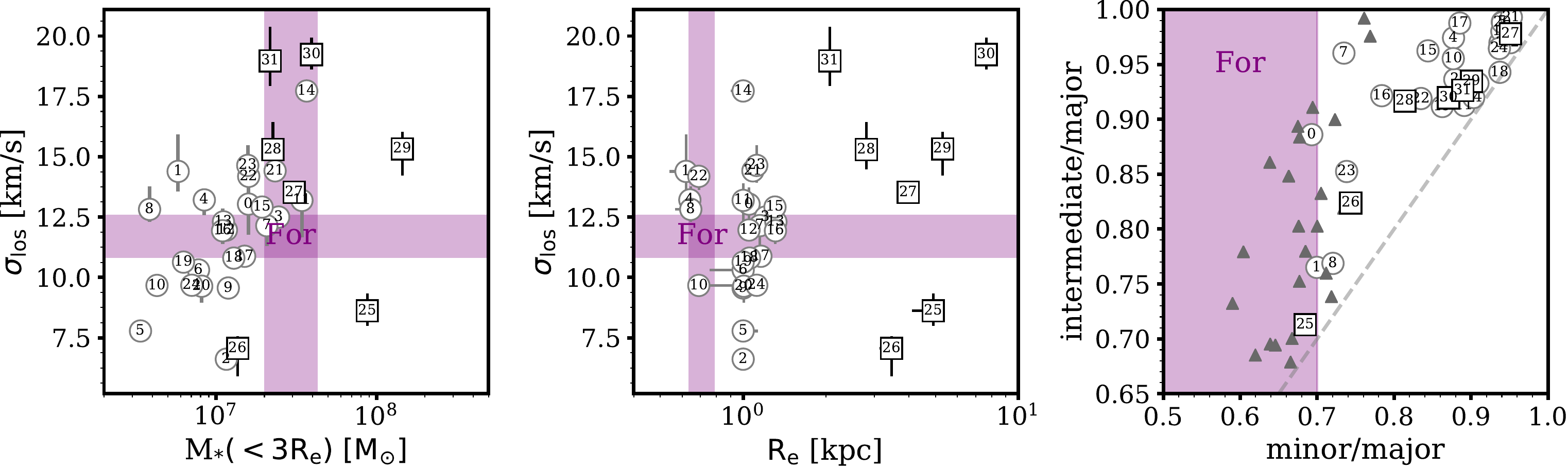}

		\caption{The sample of classical dwarf analogues. CDM dwarfs are shown with grey circles and SIDM dwarfs are shown with black squares. \textit{Left:} line-of-sight velocity dispersion as a function of stellar mass within three times the projected half-light radius. The purple bands show the measurements for Fornax from \citet{mcconnachieCensus} and \citet{fornaxdeboer}. \textit{Middle:} line-of-sight velocity dispersion as a function of the projected half-light radius. \textit{Right:} ratio of intermediate to major axes as a function of ratio of minor to major axes for our sample of dwarfs. The limits on the sphericity of Fornax dwarf galaxy (minor-to-major axis ratio $\simeq 0.7$), which has a 2D ellipticity $e \simeq 0.3$, are shown with a purple shaded band. The grey triangles show a sample of 24 isolated APOSTLE dwarfs with axis ratios comparable to the upper limit set on Fornax. The dashed one-to-one line highlights prolate galaxy shapes.}

		\label{fig1}
	\end{figure*}

\subsection{Sample of dwarfs}

For our sample of galaxies we have selected simulated dwarfs with comparable properties to classical Milky Way satellites and at least 400 bound stellar particles. We restrict our sample to satellites only, defined as the objects within 300~kpc of the Milky Way and Andromeda analogues. As the {\sc GravSphere} method relies on the standard assumptions of the spherical Jeans equation (namely, spherical symmetry, lack of rotation and equilibrium), we further restrict our sample of galaxies to those with no bound gas and those that do not exhibit significant signs of rotation. We do, however, include aspherical dwarfs in our sample. These conditions are satisfied by classical Milky Way dwarfs, for which there are no H$\rm{I}$ detections, only weak rotation is observed and asphericity is often present \citep{battagliasculptor,mcconnachieCensus}. We quantify rotation by the fraction of stellar particles which are rotating in the same direction as the total stellar angular momentum vector, $f_{\rm{corot}}$. We select galaxies with $f_{\rm{corot}}<0.6$. Note that, in order to obtain a statistically significant sample of dwarfs in both CDM and SIDM dark matter models, we do not include restrictions on the orbits or star formation histories in the selection of our sample of simulated dwarfs.

Our selection criteria have significantly cut down the available sample of high-resolution SIDM dwarfs (which have large dark matter cores), prompting an inclusion of simulated dwarfs from the lower-resolution SIDM version of the volume presented in \citet{alejandrocores}. The relative lack of suitable dwarfs in SIDM is intriguing, however the peculiarities of galaxy formation and evolution in alternative dark matter models are beyond the focus of this work (see \citealt{lovell} for details of star formation in the APOSTLE SIDM volume). 

The resulting sample consists of dwarfs with properties that are generally similar to Fornax. This includes 25 CDM dwarfs and 7 SIDM dwarfs (2 high-resolution and 5 lower-resolution). Structural and kinematic properties of the sample are plotted in Fig.~\ref{fig1}. For each dwarf in Fig.~\ref{fig1}, we show the distribution of projected half-light radii, $R_e$, from various line-of-sight projections, line-of-sight velocity dispersion, $\sigma_{\rm los}$, and stellar mass within $3R_e$, $M_*(<3R_e)$. These properties are computed using 192 isotropically distributed lines of sight generated with the {\sc healpix} algorithm \citep{healpix}. {\sc healpix} provides a convenient way of generating an isotropically distributed set of lines of sight with a more uniform distribution than one generated randomly. To calculate $R_e$, to each projection we fit a 3-component 2D Plummer profile (more details may be found in Section~\ref{allequations}) and compute the radius which contains half the projected stellar mass. The means and standard deviations of these values are shown with circles and their error bars, with errors primarily reflecting the asphericity of each system. Similarly, we compute the mass-weighted mean velocity dispersion along each projection, taking into account the error from sample size (the standard error on the mean). The stellar masses within 3$R_e$ are computed by summing the masses of stellar particles identified as bound by the {\sc subfind} algorithm \citep{subfind,dolagSubfind} and removing the contaminant stars that belong to the host galaxy\footnote{Due to the specifics of particle assignment to haloes in {\sc subfind}, some low-velocity host halo stars may end up attributed to a subhalo. We remove these stars by ensuring they were not `bound' to any given subhalo in the previous simulation snapshot \citep{subfind}.}. The contaminant stars are also removed in the calculation of other galaxy properties and the Jeans analysis. Purple bands in Fig.~\ref{fig1} show corresponding properties of the Fornax dwarf galaxy. The velocity dispersion and the half-light radius of Fornax were taken from \citet{mcconnachieCensus}. Stellar masses for Fornax span the range of values from literature \citep{mcconnachieCensus,fornaxdeboer}.

The rightmost panel of Fig.~\ref{fig1} shows intermediate-to-major axis ratios as a function of minor-to-major axis ratios, computed for the stellar component. The axes have been derived from the eigenvalues of the reduced inertia tensor, computed for the stellar particles (see e.g. \citealt{sphericity}). The purple dashed line is the upper limit set on the minor-to-major axis ratio of Fornax, determined by the measured projected ellipticity, $e = 1-b/a \simeq 0.3$ \citep{battagliafornax}. Our sample, which has an average axis ratio of $c/a\sim0.9$, is considerably more spherical than Fornax ($c/a \simeq 0.7$). This is a consequence of our selection criteria, specifically the lack of rotation and the lack of gas for the selected sample of dwarfs. These are more likely to be features of dwarf galaxies that have undergone tidal effects, which tends to reduce asphericity \citep{spherical_barber}. 

In order to further investigate the effects of asphericity we included a sample of 24 isolated CDM dwarfs with $c/a \sim 0.7$ (grey triangles in Fig.~\ref{fig1}). The sample has been chosen to contain galaxies where gas does not dominate by mass within the 3D half-mass radius of the stars. This is the reason for the lack of isolated dwarfs available for the SIDM sample. In fact, we did find 4 isolated SIDM dwarfs that match these criteria, but they turned out to be very oblate. We excluded these dwarfs from our analysis. Since the {\sc GravSphere} method explicitly accounts for the mass contributed by stars, but not by gas, when comparing {\sc GravSphere}'s performance for these dwarfs with the `true' values, we compare to the combined mass in gas and dark matter.

In order to generate the photometric and kinematic data required by {\sc GravSphere}, we obtained the stellar positions and velocities for particles in each subhalo, which were classified as bound by {\sc subfind} \citep{dolagSubfind}. We removed all contaminant stars belonging to the host halo. Each stellar system was centred at the central peak of the density field, computed using the `shrinking spheres' algorithm \citep{power}. The positions and velocities were then projected along three different lines of sight -- major, minor or intermediate axes of the subhalos. The stellar particles within 2$R_{\rm{2D}}$, where $R_{\rm{2D}}$ is the projected radius that contains half the stellar particle mass, were randomly sampled, providing 400 - 2500 particles for the kinematic sample of each dwarf. The photometric sample included 400 - 2500 particles and was chosen to be the same or bigger in size than the kinematic sample. The velocities within the kinematic sample were perturbed with Gaussian noise with a standard deviation of 2~km~s$^{-1}$, representing typical measurement errors. The mass within $3R_{\rm 2D}$ has additionally been provided, $M(<3R_{2D})$, such that {\sc GravSphere} may include the stellar mass contribution to the gravitational potential in the Jeans modelling.

\subsection{The {\sc pyGravSphere} code}

In this work we present a {\sc Python} implementation of the {\sc GravSphere} method, {\sc pyGravSphere}. {\sc pyGravSphere} is open source software\footnote{\url{https://github.com/AnnaGenina/pyGravSphere}}. As in the work of \citet{gravsphere}, {\sc pyGravSphere} is based on the affine-invariant ensemble sampler {\sc emcee} \citep{emcee}. {\sc emcee} differs from the classic Metropolis-Hasting algorithm in that each individual Markov chain, or `walker', communicates with the other `walkers' at each step, thus allowing the chains to efficiently sample the posterior distribution. {\sc emcee} has parallel functionality, which we exploit in this work. In the following, we outline the assumptions and parameters that enter into our {\sc emcee} setup.

\subsubsection{{\sc emcee} parameters}

\label{allequations}

As in \citet{gravsphere}, to parametrize the dark matter distribution, {\sc pyGravSphere} employs a broken power-law model with 5 spatial bins defined as logarithmically spaced fractions of the half-light radius\footnote{In this work we use multiple definitions of the half-mass radius, which we list here for the purposes of clarification. $R_{2D}$ is the projected radius which contains half of a dwarf's stellar mass and it is computed by direct summation. $R_e$ is the projected radius containing half the stellar mass, derived from a 3-component Plummer profile fit to the \textit{sample} of stellar particle data. We frequently refer to $R_e$ as the half-light radius.}, $R_e$, with bins $r_j = \text{[0.25, 0.5 ,1, 2, 4] }R_e$. Within each bin, the density follows a power law defined by slopes $\gamma_j$. The overall distribution is described by  

\begin{equation}
    \centering
    \rho_{\rm{dm}}(r) = 
    \begin{cases}
    \displaystyle \rho_{0} \left(\frac{r}{r_0} \right)^{-\gamma_0}, &  r < r_0 \\ 
    \displaystyle \rho_{0}\left(\frac{r}{r_{j+1}} \right)^{-{\gamma_{j+1}}} \prod_{n=0}^{n<j+1} \left(\frac{r_{n+1}}{r_n} \right)^{-{\gamma_{n+1}}},  &   r_j < r < r_{j+1} \\
   
    \end{cases}
 \end{equation}
where $\rho_0$ is the density at $r_0$. Beyond the outermost bin, the power law is extrapolated. Note that this radial extent typically covers the positions of available kinematic tracers. 

For the light profile, {\sc {pyGravSphere}} uses a sum of three \citet{plummer} components ($N_P = 3$):

\begin{equation}
\nu(r) = \displaystyle \sum_{j}^{N_{P}} \frac{3M_j}{4\pi a^3_j}\left( 1 + \frac{r^2}{a^2_j}\right)^{-5/2},
\end{equation}
where $M_j$ and $a_j$ are the relative weight and spatial extent of each component, respectively. This distribution is straightforward to project, yielding:

\begin{equation}
\label{plummereq}
\Sigma(R) = \displaystyle \sum_{j}^{N_P} \frac{M_j}{\pi a_j^2}\left( 1+\frac{R^2}{a_j^2}\right)^{-2}.    
\end{equation}
The velocity anisotropy is parametrized following \citet{baes}:

\begin{equation}
\centering
   \displaystyle \beta(r)=\beta_0+(\beta_{\infty} - \beta_0)\frac{1}{1+\left(\frac{r_t}{r} \right)^\eta},
\end{equation}
where $\beta_0$ is the central value of the anisotropy, $\beta_{\infty}$ is the value at infinity, $r_t$ is the radius of transition and $\eta$ is its steepness. 

\subsubsection{{\sc pyGravSphere} data input}
The photometric sample of stars is split into bins of $N_{\rm phot}/\sqrt{N_{\rm phot}}$ particles per bin, where $N_{\rm phot}$ is the size of the photometric sample. This choice allows for efficient spatial coverage and low Poisson error. We weight each particle by the relative number of stars it represents (i.e. we define the weight of each particle as $w_{\rm p} = m_{\rm p}N_{\rm tot}/M(<3R_{\rm 2D})$, where $N_{\rm tot}$ is the total number of particles in the sample). For each photometric bin we calculate the Poisson errors and use the {\sc lmfit} algorithm \citep{lmfit} to obtain the best 3-component Plummer fit (Equation~\ref{plummereq}). This profile is then input into {\sc emcee}. 

The kinematic data are also split into bins of $N_{\rm kin}/\sqrt{\rm N_{kin}}$ particles per bin, where $N_{\rm kin}$ is the size of the kinematic sample. The error in each bin is computed by adding the Poisson and sampling errors in quadrature, where we again weight each particle by $w_p$. This procedure is described in detail in \citet{gravsphere}. The effect of the number of particles per bin was explored in \citet{gravsphere}, who found little impact on their results. We confirm this to be the case, provided the signal-to-noise is not low due to `overbinning' in the inner, dense, regions.

We use the same kinematic bins to compute the mean and errors of the two VSPs. Because VSP2 is sensitive to the behaviour of the velocity dispersion profile in the outer regions (due to the $R^3$ term), we fit a power law to the computed $v^4_{P}$ profile outside of the projected half-light radius and extrapolate it following \citet{readdraco}.

\subsubsection{{\sc emcee} set-up and priors}

\begin{table}
\centering
\caption{Default {\sc GravSphere} priors}
\begin{tabular}{l|c|c}
Property & Parameter & Prior \\
\hline \hline
Dark Matter & $ \log_{10}\rho_0$/M$_{\odot}$kpc$^{-3}$   & [5, 10] \\
& $\gamma_{0,1,2,3,4}$    &  [0, 3], $\Delta \gamma_{\mathrm{max}}=1$\\
\hline
Anisotropy & $\tilde{\beta}_0$ & [-1, 1]\\
&$\tilde{\beta}_{\infty}$ & [-1, 1]\\
&$\log_{10}r_t$/kpc & $\log_{10}$[$R_e/2$, $2R_e$]\\
&$\eta$ & [1, 3] \\
\hline
Tracers & $\log_{10}M_j$/M$_{\odot}$ & $\log_{10}$[$M_{\rm bf,j}/2$ { ,   }$3/2M_{\rm bf,j}$ ] \\
& $a_j$/kpc & [$a_{\rm bf,j}/2${   ,   } 3/2$a_{\rm bf,j}$] \\
\hline
Baryons & $\log_{10}M_*$/M$_{\odot}$ & $\log_{10}$[$0.75M(<3R_{\rm 2D})$, \\
& & $1.25M(<3R_{\rm 2D})$] 

\end{tabular}
\label{table1}
\end{table}

The priors on each of the parameters in the default {\sc pyGravSphere} set-up are shown in Table~\ref{table1}. Parameters $\log_{10} M_j$ and $a_j$ of the best 3-component Plummer fit are allowed to vary within 50 per cent of their linear best-fit values, as determined by {\sc lmfit}, while the stellar masses $\log_{10}M_*$ were varied within 25 per cent of the $M(<3R_{2D})$ value. 

To ensure that the walkers sample the whole hypervolume of parameter space, defined by the parameter constraints, the starting positions of walkers are ideally generated to follow a uniform distribution. This is difficult to achieve when the values of the dark matter slope, $\gamma_j$, in each radial bin, $r_j$, are constrained to monotonically increase. We find that typically $\sim$ 5 per cent of the initially generated walker positions fall within these defined bounds. If the chains are allowed to run for long enough, the walkers that are `stuck' in forbidden regions of the parameter space may eventually make their way to the allowed regions. This process is, however, dependent on the efficiency of the active walkers in probing the posterior distribution. In cases where the posterior is multimodal, for example, some regions of the parameter space will not be probed due to the nature of the ensemble sampler, where the walkers communicate with each other, unlike in the classical Metropolis-Hastings algorithm.

We thus use the following procedure to generate the initial walker positions. For each walker, we generate the free parameters following a uniform distribution. We then throw away the walkers that do not satisfy our $\gamma_j$ constraints. For the discarded walker positions we generate new ones, accepting those that satisfy the constraints and rejecting the others. This procedure is repeated recursively until each walker has a randomly generated initial position that satisfies the constraints. The effective priors and the advantages of the method over the one implemented in the original works using {\sc GravSphere} (e.g. \citealt{gravsphere,readdraco}) are discussed in Appendix~\ref{convergence}.

We use 1000 walkers to probe the posterior distribution. Each walker is run for $10^4$ steps as a conservative `burn-in' measure, and then for a further $10^4$ steps. The results presented in Appendix~\ref{convergence} suggest that our walkers are converged after $\sim6\times10^3$ steps, so the above choices are rather extreme.

Since the anisotropy parameter, $\beta$, can take on values between 1 and $-\infty$, one would benefit from transforming this into a finite range. As in \citet{gravsphere}, we use the symmetrized anisotropy parameter,

\begin{equation}
    \tilde{\beta} = \frac{\beta}{2-\beta}, \quad {\rm with ~~~} {-1<\tilde{\beta}<1}.
\end{equation}

This symmetrized form of the anisotropy allows the {\sc emcee} walkers to probe the entire range of possible anisotropy values. In practice, we apply the constraints $\tilde\beta_0 > -0.95$ and $\tilde\beta_{\infty} > -0.95$, as for more negative values the calculation becomes numerically unstable.

{\sc pyGravSphere} solves the Jeans equation for the projected velocity dispersion profile, $\sigma_P(R)$. It additionally fits the projected number density distribution, $\Sigma(R)$, and the two VSPs. We define the `log-likelihood function' as the chi-squared sum of these four components:

\begin{equation}
\label{likelihood}
\displaystyle \ln{\mathcal{L}} = -\frac{1}{2}\left(\chi^2_{\sigma_{LOS}} + \chi^2_{\Sigma} + \chi^2_{VSP1} + \chi^2_{VSP2} \right)  .
\end{equation}

We note that Equation~\ref{likelihood} involves quantities that are, to some extent, correlated. We must therefore consider whether the form of the likelihood function in Equation~\ref{likelihood} is justified. It has been pointed out in \citet{lokasmamoncoma} and \citet{lokas2009} that the correlation between the second and the fourth velocity moments is typically weak. Aside from the fourth velocity moment, the virial shape parameters are also related to the projected surface density $\Sigma(R)$. Given that the photometric samples for classical dwarfs are typically large, $\Sigma(R)$ is measured to a sufficiently high accuracy, such that its uncertainties are negligible compared to those of the fourth velocity moment. Moreover, \citet{gravsphere} find that the estimates of the second velocity moment and the virial shape parameters are typically normally distributed. This, together with the weak correlation between the second and fourth velocity moments, suggests that Equation~\ref{likelihood} is a good approximation to the true likelihood function. 

\begin{figure*}

		\includegraphics[width=1.7\columnwidth]{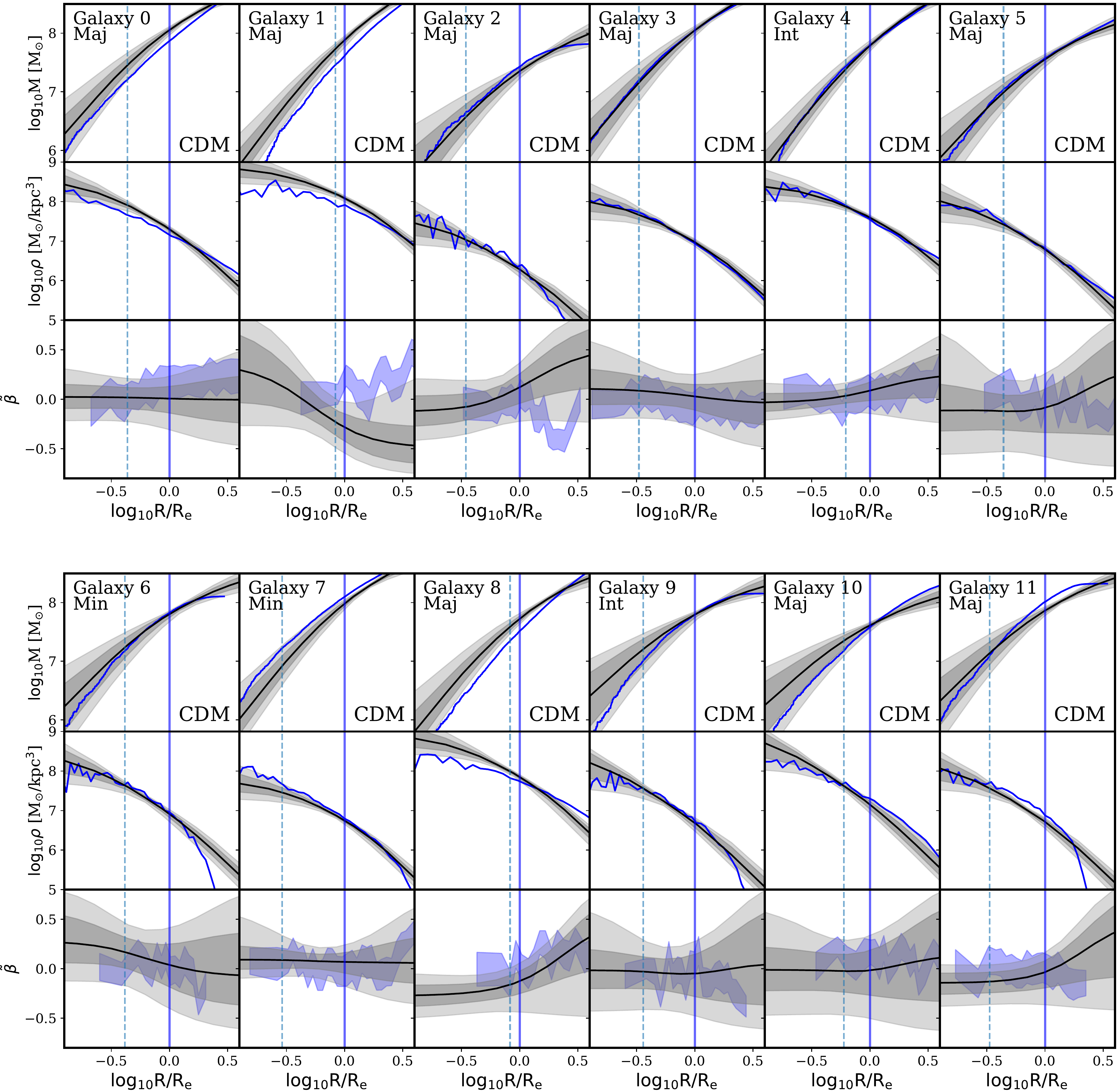}

		\caption{ Enclosed dynamical mass profiles (top), density profiles (middle) and symmetrized anisotropy profiles recovered for a sample of 32 simulated dwarf galaxies using the {\sc GravSphere} method. Only one projection (from the centre of the host galaxy) is show for each dwarf. The profiles are shown as a function of 3D radius, normalized by the projected half-light radius, $R_e$. The black lines and the dark and light grey shaded bands display the median, 68 and 95 per cent confidence limits, respectively. The `true' mass and density profiles, measured directly from the simulation, are shown in blue. The solid vertical blue line shows the location of the projected half-light radius and the dashed blue line shows 2.8$\epsilon$ ($\approx$380~pc for CDM dwarfs and SIDM dwarfs (25,26) and $\approx$655~kpc for SIDM dwarfs 27-31), which is close to the convergence radius for these systems. The `true' velocity anisotropy, as measured directly from the stellar particles, is shown with a shaded blue band. For each galaxy, in the top left corner, we display the unique galaxy number and the principal axis along which the galaxy was projected to produce this figure (which is the principal axis most closely aligned with the vector to the host galaxy). In the bottom right corner we show whether the galaxy is from CDM or SIDM cosmology.}

		\label{fig2}
	\end{figure*}
 
	\begin{figure*}

		\includegraphics[width=1.7\columnwidth]{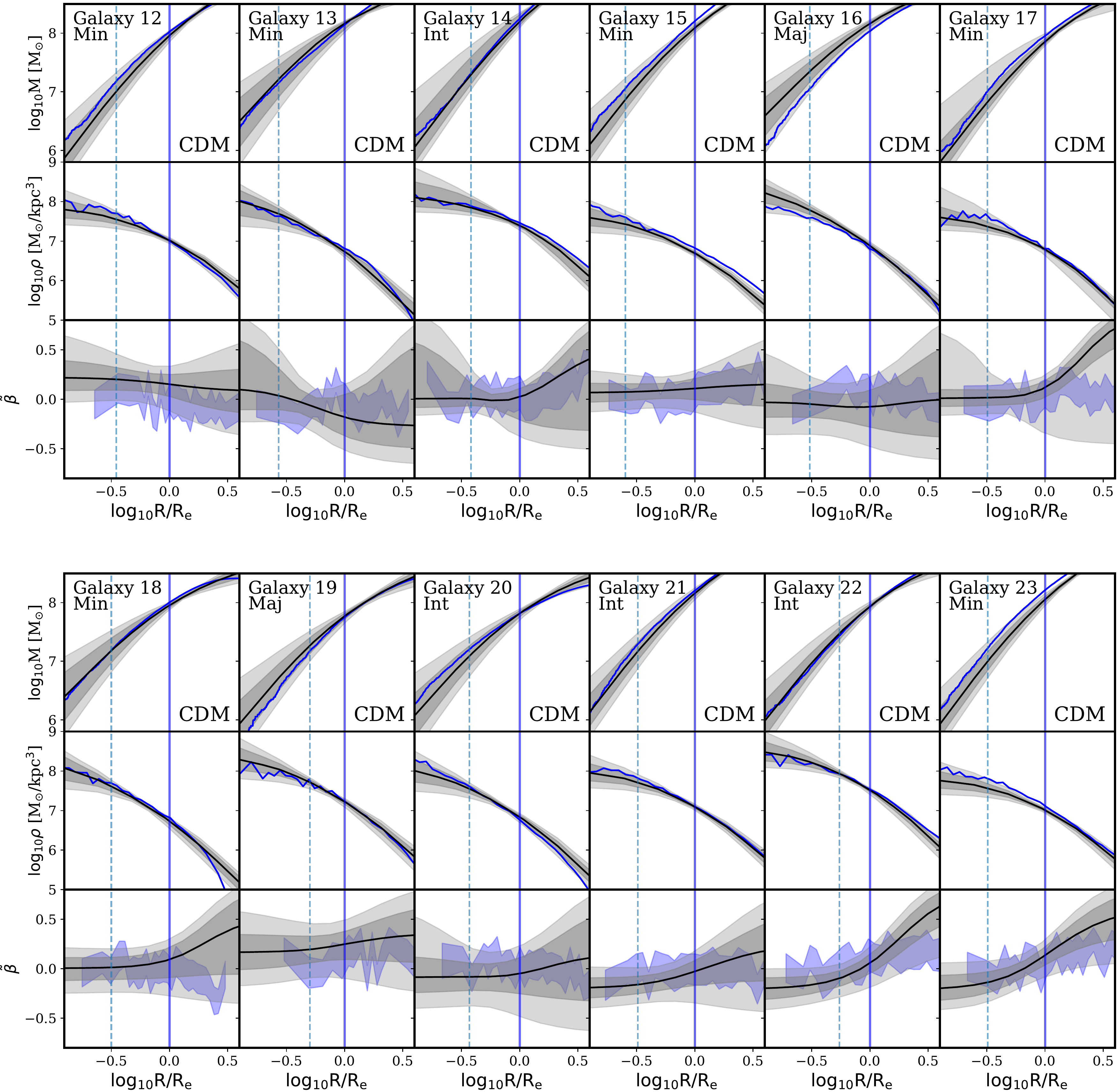}

		\contcaption{ }

		\label{fig2a}
	\end{figure*}
	
	\begin{figure}

		\includegraphics[ width =\columnwidth]{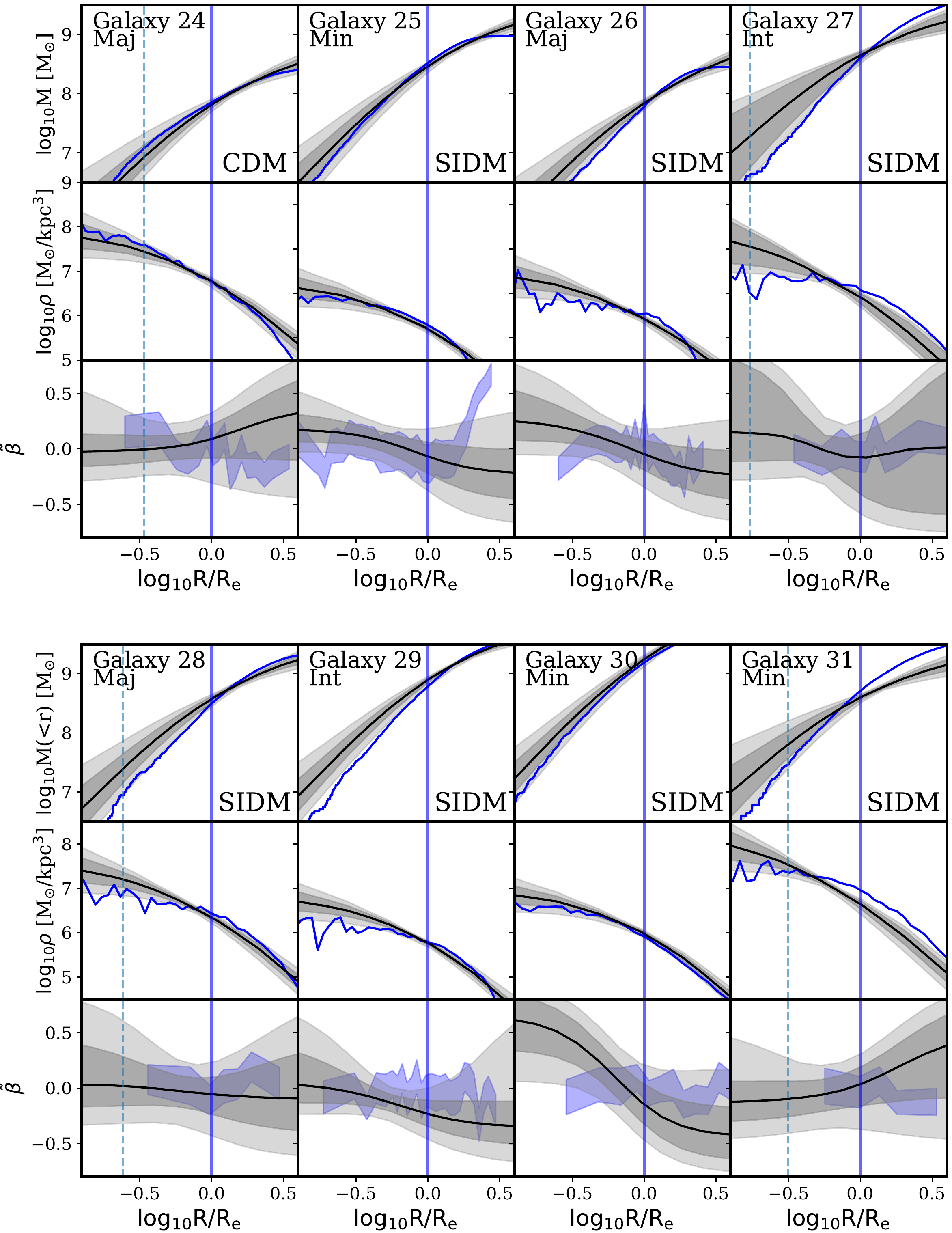}

		\contcaption{Galaxies 25 - 31 are SIDM dwarfs. Note that for a number of these dwarfs the spatial resolution, $2.8\epsilon$ (vertical dashed line), is below 0.125$R_e$ and below the limits of the figure. }

		\label{fig2b}
	\end{figure}

\begin{figure*}{
   \begin{multicols}{2}
	\includegraphics[width=\columnwidth]{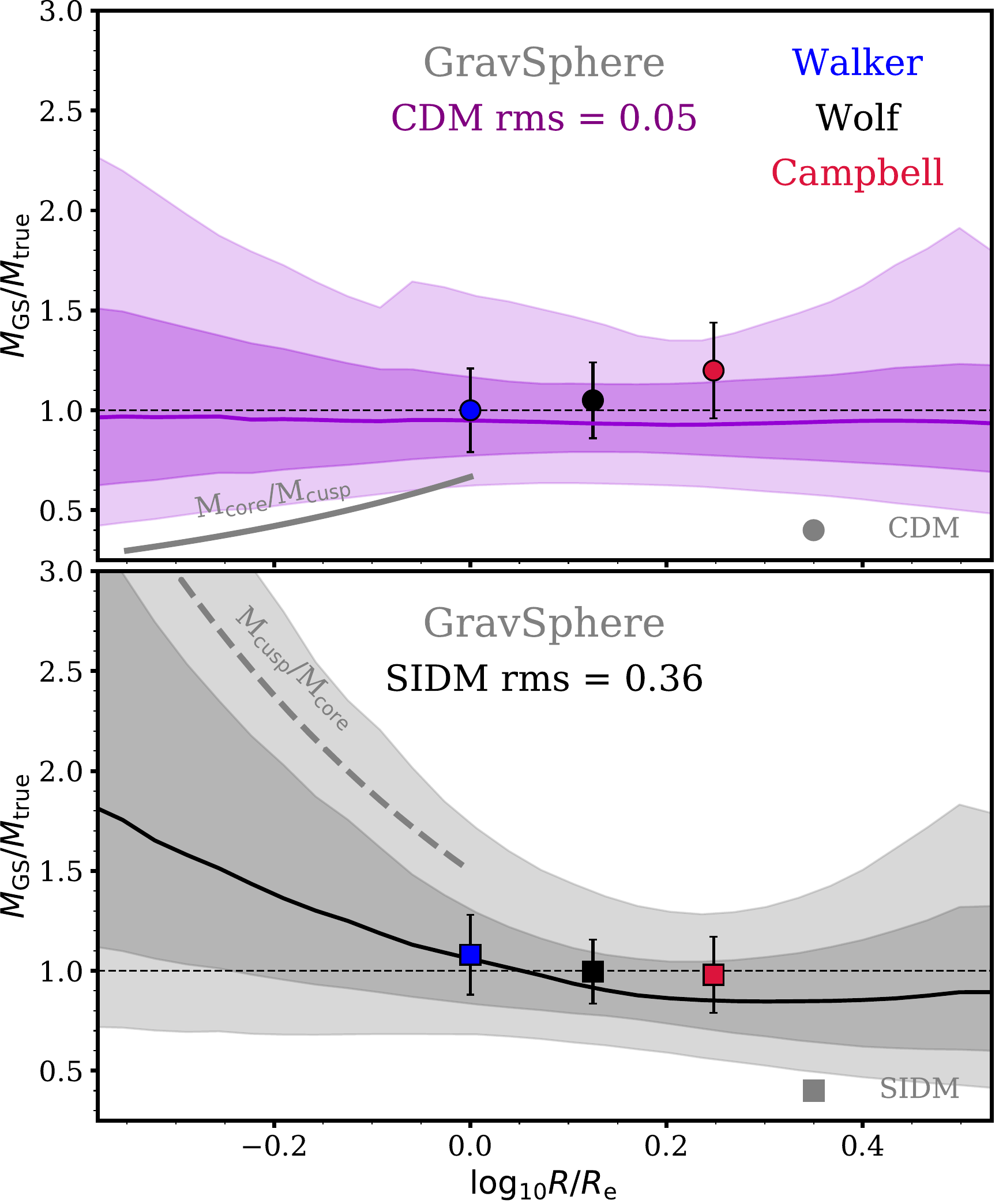}\par
    \includegraphics[width=\columnwidth]{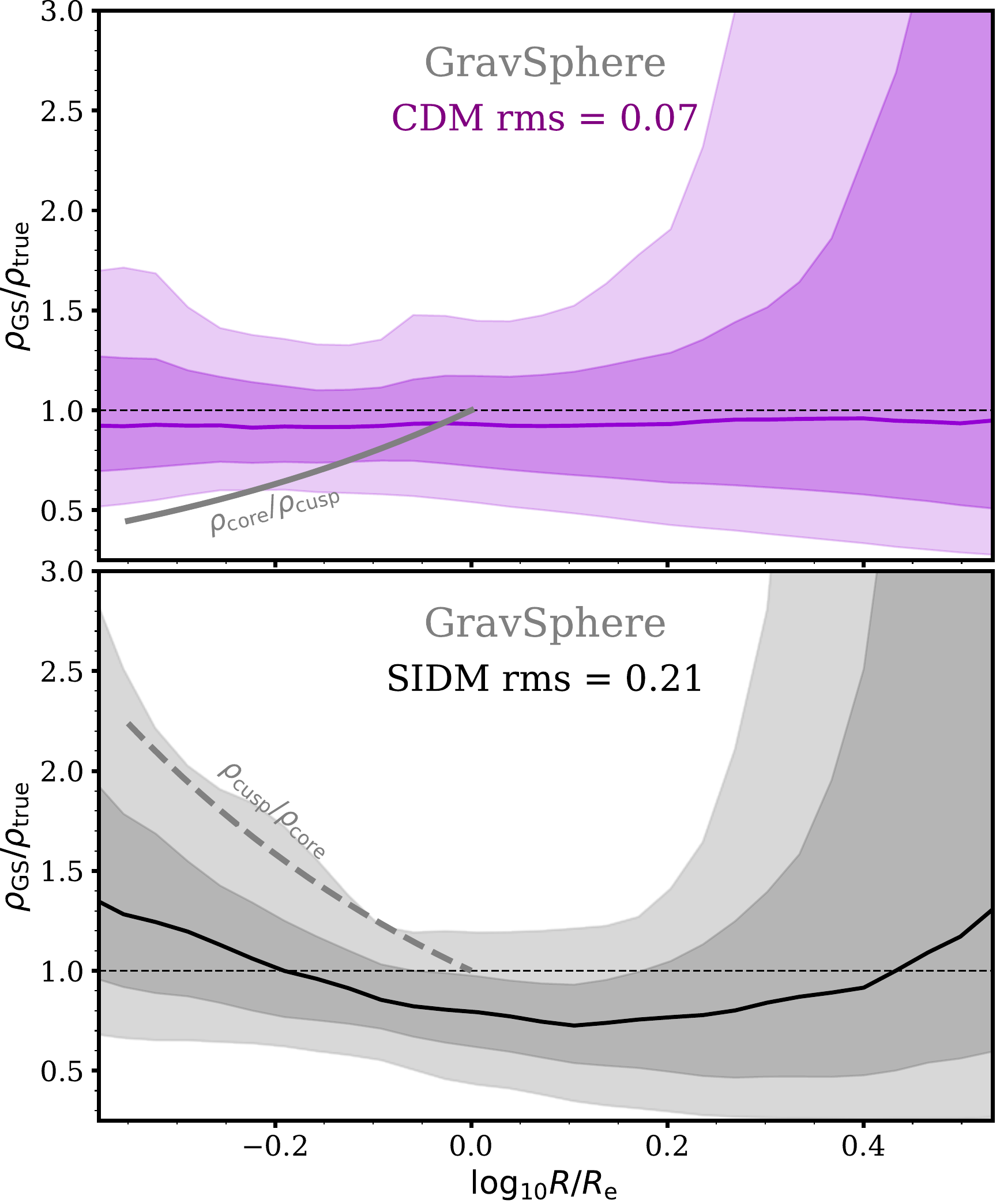}\par
    \end{multicols} 
      
		\caption{\textit{Left:} ratio of the recovered to true enclosed mass as a function of radial distance in units of the half-light radius, $R_e$. The purple line and purple shaded bands show the median, 68 and 95 per cent confidence limits for the CDM galaxy sample. The black line and the corresponding shaded bands show the sample of SIDM dwarfs. The symbols and their error bars show the accuracy of standard estimators from \citet{walkerest}, \citet{wolf} and \citet{campbell}, identified by their colours. Circles are for the CDM dwarfs and squares for SIDM. The `rms' values show the root-mean-square offset of the median value from 1 along the entire radial range displayed. The grey dashed line shows the radially-dependent  bias in the mass profiles that is required to infer an NFW cusp, when in reality there is a core. The solid grey line shows the bias required to infer a dark matter core, when in reality there is a cusp. \textit{Right:} ratio of the recovered to true density as a function of radial distance in units of the half-light radius. The grey dashed line shows the bias in the recovered density profiles required to infer an NFW cusp, when in reality there is a core. The solid grey line shows the corresponding bias for an incorrect inference of a core, when the true profiles are cuspy.}

		\label{fig3} }
\end{figure*}

\section{Results}
\label{comparison}
In this section we present the performance of the {\sc GravSphere} method, under a default setup, on each of the dwarfs in our sample. We further quantify its global performance, for the entire sample, and compare to standard Jeans analysis approaches. We identify the causes of bias and scatter in the recovered enclosed mass profiles.

\subsection{Individual dwarfs}

Fig.~\ref{fig2} shows the recovery by {\sc GravSphere} of the cumulative mass profiles, density profiles and the velocity anisotropy profiles of all dwarfs in our sample. In this figure we only display results for projections along the vector to the centre of the host galaxy. The `true' mass, density and anisotropy profiles are displayed in blue and the {\sc GravSphere} results in grey shaded bands. To generate the {\sc GravSphere} profiles we took 10$^5$ random samples from the output MCMC chains and for each radial position computed the median and the 68$^{\rm th}$ and the 95$^{\rm th}$ percentiles. The choice of 10$^5$ samples is sufficient to produce representative posteriors, but is otherwise arbitrary.

Mass profiles of simulated dwarfs were computed by summing dark matter particle masses radially from the centre of each dwarf, where we define the centre as the centre of mass of the stellar component. Densities were computed in 31 logarithmically-spaced bins in radius, from $\log_{10}r/{\rm kpc} = -2$ up to the furthest bound dark matter particle. In order to compute the stellar velocity anisotropy profiles of simulated dwarfs, we bin the stars in each dwarf into 50 logarithmically spaced bins, starting from the position of the star that is closest to the centre of mass and ending at the outermost star. We then reduce the number of bins and widen the bin edges such that each bin has at least 50 stars within. We construct 1$\sigma$ error bars by taking 1000 random samples of 25 stars with replacement and computing the standard deviation of the velocity anisotropy in each bin.

In this analysis we focus on the key region within the half-light radius of each galaxy (solid vertical blue lines in Fig.~\ref{fig2}) and above $2.8\epsilon$ (dashed blue lines). {\sc GravSphere} appears to be unbiased on average. For the CDM sample, the dwarf mass, density and anisotropy profiles are typically contained within the 68 per cent confidence limits, although the uncertainties can be large. Notable exceptions in the CDM sample are Galaxies 1 and 8. These are known aspherical objects (see Fig.~\ref{fig1}). Similar systematics can be seen for the remainder of the aspherical sample of dwarfs. It is clear that in SIDM dwarfs the enclosed mass is {\it always} overestimated, although even in these galaxies the true profiles are contained within the 95 per cent confidence regions. We will return to this issue in Section~\ref{sidmsec}.

We also note that the symmetrized anisotropy profiles in our simulations are generally consistent with being constant and isotropic ($\tilde{\beta}=0$). Deviations from this are seen in galaxies that are aspherical, and, as we will see in Section~\ref{tidessection}, in those which are affected by tides.

\subsection{Comparison to standard estimators}
We now compare the performance of {\sc GravSphere} in recovering enclosed masses to three mass estimators from the literature. The mass estimators take the form:
\begin{equation}
M(<\mu R_e) = \frac{\lambda R_e \langle\sigma_{P}^2\rangle}{G},
\end{equation}
where $\mu$ and $\lambda$ are constants. 

The estimator provided by \citet{wolf} gives the mass enclosed within the deprojected half-light radius $R_3$, with $R_3$ $\approx 4/3R_e$; the \citet{walkerest} estimator gives the enclosed mass at the projected half-light radius, $R_e$. Another estimator has been derived by \citet{campbell} for the mass within 1.44R$_{\rm 2D}$, where R$_{\rm 2D}$ is the projected radius containing half the stellar mass. The latter has been calibrated on dwarfs from the APOSTLE suite of simulations that we use here. We omit the estimator derived by \citet{errani}, which we find produces similar results to those of \citet{campbell}. 

As input for the estimators we use the half-light radius, $R_e$, obtained from the best-fit 3-component Plummer profile, as this is expected to provide more accurate results than the circular radius containing half the stellar mass, $R_{2D}$ \citep{fireest}. The latter definition was used in the calibration of the \citet{campbell} estimator, however we refrain from using this definition as the spatial extent of the sample of stellar particles in each dwarf is already cut off at $2R_{2D}$. For \citet{wolf} and \citet{walkerest} estimators we calculate the mean velocity dispersion and the associated errors using the technique of \citet{walkerfornax}, where we incorporate weighting by the number of stars per stellar particle, $w_p$, in the likelihood function. For the \citet{campbell} estimator we compute the mass-weighted mean velocity dispersion below 1.04$R_e$, as prescribed.

In the left panel of Fig.~\ref{fig3}, we show the bias in mass profiles returned by {\sc GravSphere} (mass recovered by {\sc GravSphere} divided by the true mass) and the 68 and 95 per cent confidence levels, as a function of normalised radius, $R/R_e$. We use the following procedure to compute the global radial bias and the associated confidence levels. For each galaxy and for each distance, $R$, we obtain a cumulative mass distribution from 10$^5$ random samples of the MCMC posteriors. We then use this cumulative distribution for Monte Carlo sampling of accuracies, combining the samples from all galaxies at each distance. This allows us to take into account the asymmetry in the {\sc GravSphere} confidence limits, as seen in Fig.~\ref{fig2}.

The lower axis limit on $R/R_e$ in Fig.~\ref{fig3} has been chosen to be the average value of 2.8$\epsilon/R_e$ for our sample and the upper axis limit was chosen so as to contain the smallest dark matter halo in the sample. The bias and associated errors for each galaxy are only included in the making of this figure for distances $R>2.8\epsilon$. 

In purple, we show the CDM sample and in black the SIDM sample. The symbols with error bars show the performance of dynamical mass estimators from \citealt{walkerest}, \citealt{wolf} and \citealt{campbell} (circles for CDM and squares for SIDM). The \citet{wolf} and \citet{walkerest} estimators are accurate to better than 10~per~cent, however we observe a bias in the \citet{campbell} estimator. This is likely due to the aforementioned difference in the definition of the half-light radius. In all cases, the true mass is contained within the uncertainty of the estimators.

From the left panel of Fig.~\ref{fig3} it is clear that, for CDM dwarfs, {\sc GravSphere} performs just as well as the \citet{wolf} and \citet{walkerest} estimators, and with similar scatter. The inferred masses are, on average, very accurate across the entire radial range, with the scatter becoming more significant in the innermost regions as well as the outer regions. For the large values of $R/R_e$ this is caused by the lack of kinematic tracers. Moreover, our priors permit only a narrow range of slopes ($0<\gamma_j<3$), whilst our sample is expected to have undergone tidal effects, resulting in an outer slope $\gamma \approx 4$ \citep{penarrubiatides}. Nevertheless, the estimate is accurate, on average, out to $3 R_e$, with a root mean square (RMS) fractional error\footnote{We define RMS = $\sqrt{\frac{1}{N} \sum_{i = 1}^{i=N} (M_{\rm calc}/M_{\rm true} - 1})^2$, where $M_{\rm calc}$ is the mass obtained through {\sc GravSphere}, or another method, $N$ is the number of radial intervals at which $M_{\rm calc}$ is computed and $M_{\rm true}$ is the true enclosed mass at these intervals, found directly from the simulation. We use a maximum of 30 intervals to construct Fig.~\ref{fig3}.} of only 5~per~cent. It is clear that the masses for SIDM dwarfs are significantly overestimated in the inner regions (on average, up to 80 per cent), with very large scatter. The mass is accurate near the half-light radius and beyond that it is underestimated by $\sim20$~per~cent.

\subsection{The core-cusp problem}

What do these results mean in the context of the core-cusp problem? We now consider two ways to infer cores or cusps in dark matter haloes: via their characteristic inner density (as in \citealt{heating}) and via an accurate inference of profile shape.

\subsubsection{Cores vs. cusps via characteristic densities}

In the right panel of Fig.~\ref{fig3} we show the recovery by {\sc GravSphere} of the dark matter density profiles of our sample of dwarfs. For CDM dwarfs, the density profiles are accurate across the entire radial range (RMS = 0.07) and the scatter is only $\sim$30 per cent in the inner regions. In fig.~5 of \citet{heating} it can be seen that, for a dwarf of Fornax-like pre-infall halo mass, the core and cusp-like densities, $\rho_{150}$, differ by a factor of at least $3.5$. The spatial resolution of our simulations does not allow us to probe radii below 380~pc; however, if {\sc GravSphere} provides a similar level of bias and scatter for $\rho_{150}$ (corresponding to $\log_{10}R/R_e\approx-0.7$ for a Fornax-size dwarf), it is certainly possible to differentiate between the core and cusp-like densities, provided  there is complete core formation below the half-light radius and no reduction of central dark matter density due to tides \citep{alltheway}. For SIDM dwarfs, the density is overestimated in the inner regions by up to 50~per~cent and underestimated above the half-light radius by $\sim30$~per~cent, reflecting the pattern with enclosed mass. This suggests that in the case of SIDM dwarfs with cores, {\sc GravSphere} is biased towards cusps in its standard configuration. 

\subsubsection{Cores vs. cusps via profile shape}

Let us now approximate the density profiles in the inner regions by a single power law
\begin{equation}
\label{powerlawprof}
\rho(r) = \rho(R_e)\left(\frac{r}{R_e}\right)^{-\gamma},    
\end{equation}
where $\rho(R_e)$ is the density at $R_e$ and $\gamma$ is the slope of the power law, with $\gamma=0$ corresponding to a core and $\gamma=1$ to a cusp. If we assume that a core forms below the radius $R_e$, where otherwise the density within $R_e$ follows $\rho \propto r^{-1}$, then the density ratio is
\begin{equation}
\label{falsecorerho}
\frac{\rho_{\rm cusp}}{\rho_{\rm core}} = \left(\frac{r}{R_e} \right)^{-1}  ,
\end{equation}
and the mass ratio is
\begin{equation}
\label{falsecoremass}
\frac{M_{\rm cusp}}{M_{\rm core}} = \frac{3}{2}\left(\frac{r}{R_e} \right)^{-1}.
\end{equation}

We display these relations, and their inverse, with dashed (an incorrectly inferred cusp) and solid (an incorrectly inferred core) grey lines in Fig.~\ref{fig3}. The relations are displayed out to the radius $R=R_e$ for the case where a core forms below the half-light radius\footnote{We have verified that the relations in Equations~\ref{falsecorerho} and \ref{falsecoremass} are good approximations for a dwarf with a Fornax-like pre-infall halo mass \citep{heating} with full core formation below the half-light radius and an NFW profile otherwise \citep{alltheway}.}. Beyond this radius, the relations are expected to converge to 1. The relations should be taken more as visual guides for the radial dependence of the bias that we expect in order to incorrectly infer a core or a cusp. Note that for a core that forms at some fraction of the half-light radius, we would only need to scale the relation in Equation~\ref{falsecorerho} by a corresponding factor (or shift the relation up and down in $\log$-space). Unfortunately, cores on the scale of $0.5R_e$ are too small to be probed by our simulations for the dwarf galaxy masses we consider here.   

For the case of CDM dwarfs, it is clear that within the 68~per~cent confidence regions the mass and density profiles returned by {\sc GravSphere} are fully consistent with cusps. Cores lie outside the 95~per~cent confidence regions. For SIDM dwarfs there is a clear bias towards more cuspy profiles. Fully cored profiles below the half-light radius are contained within the 68~per~cent regions, but so are cusps that are only slightly shallower than $\rho \propto r^{-1}$. 

\subsection{Comparison to other methods and parametrizations}

\begin{figure*}
\setlength\multicolsep{0pt}
\begin{multicols}{2}
	\includegraphics[width=0.95\columnwidth]{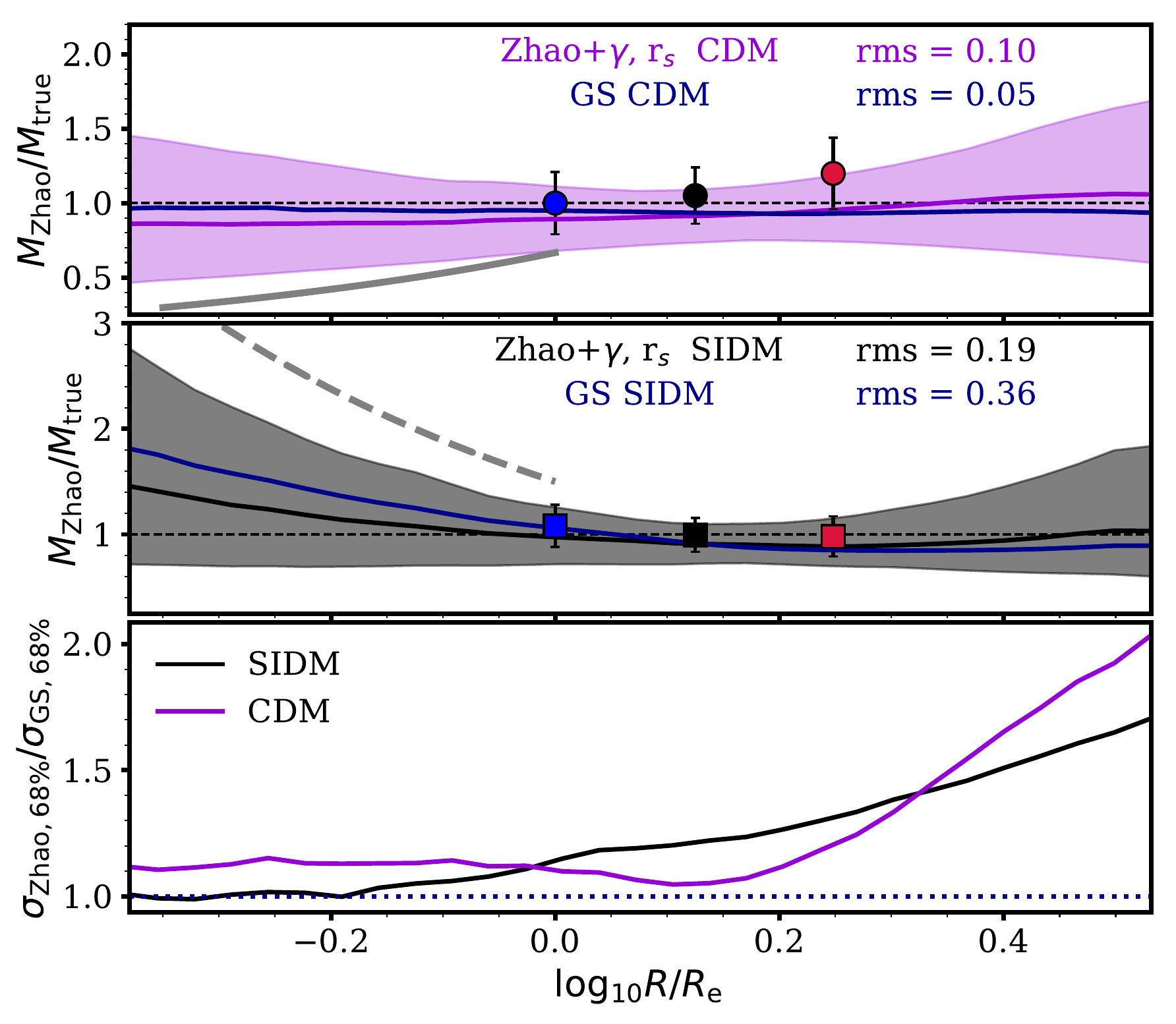}\par
    \includegraphics[width=0.95\columnwidth]{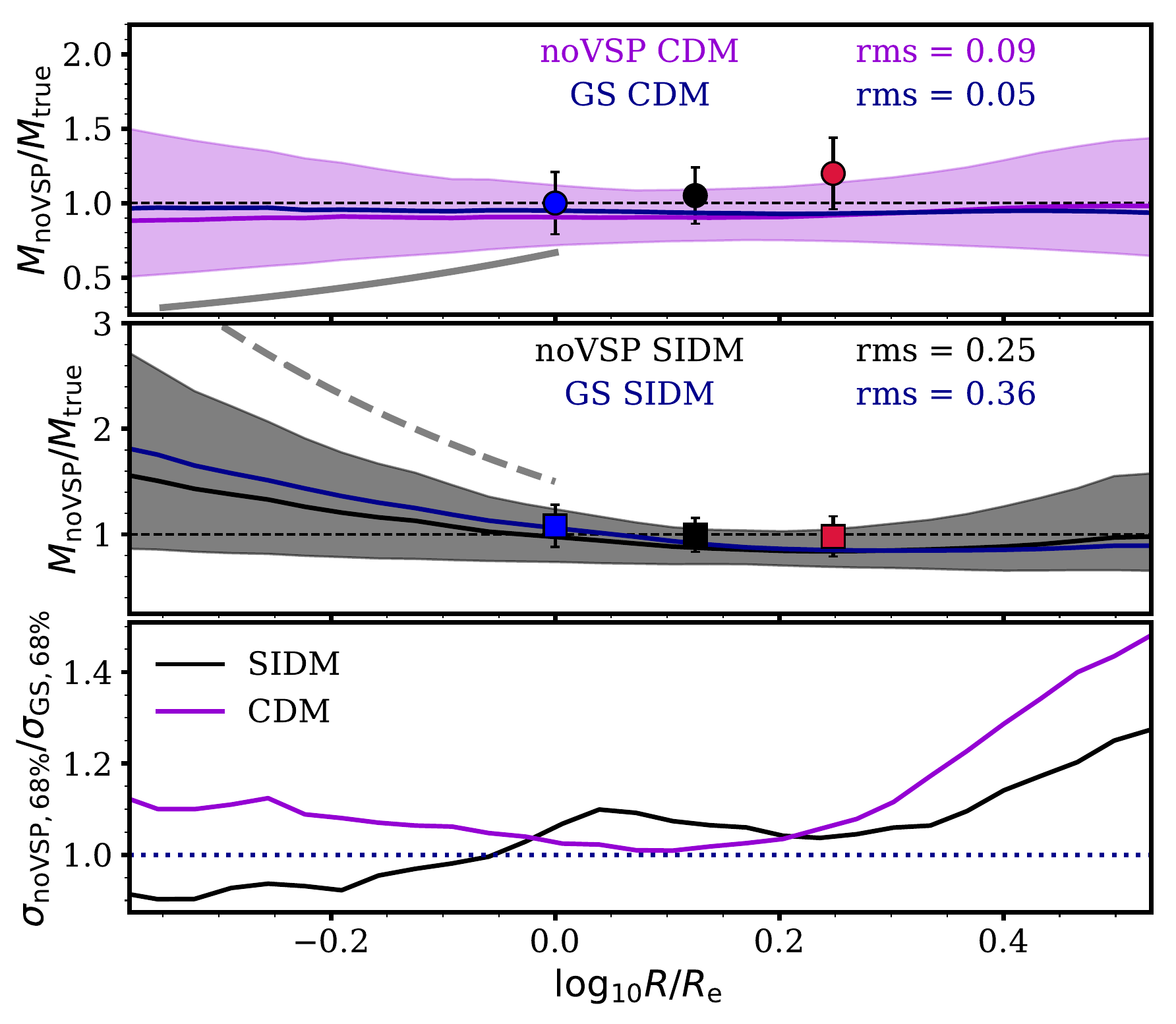}
\end{multicols}
\begin{multicols}{2}
    \includegraphics[width=0.95\columnwidth]{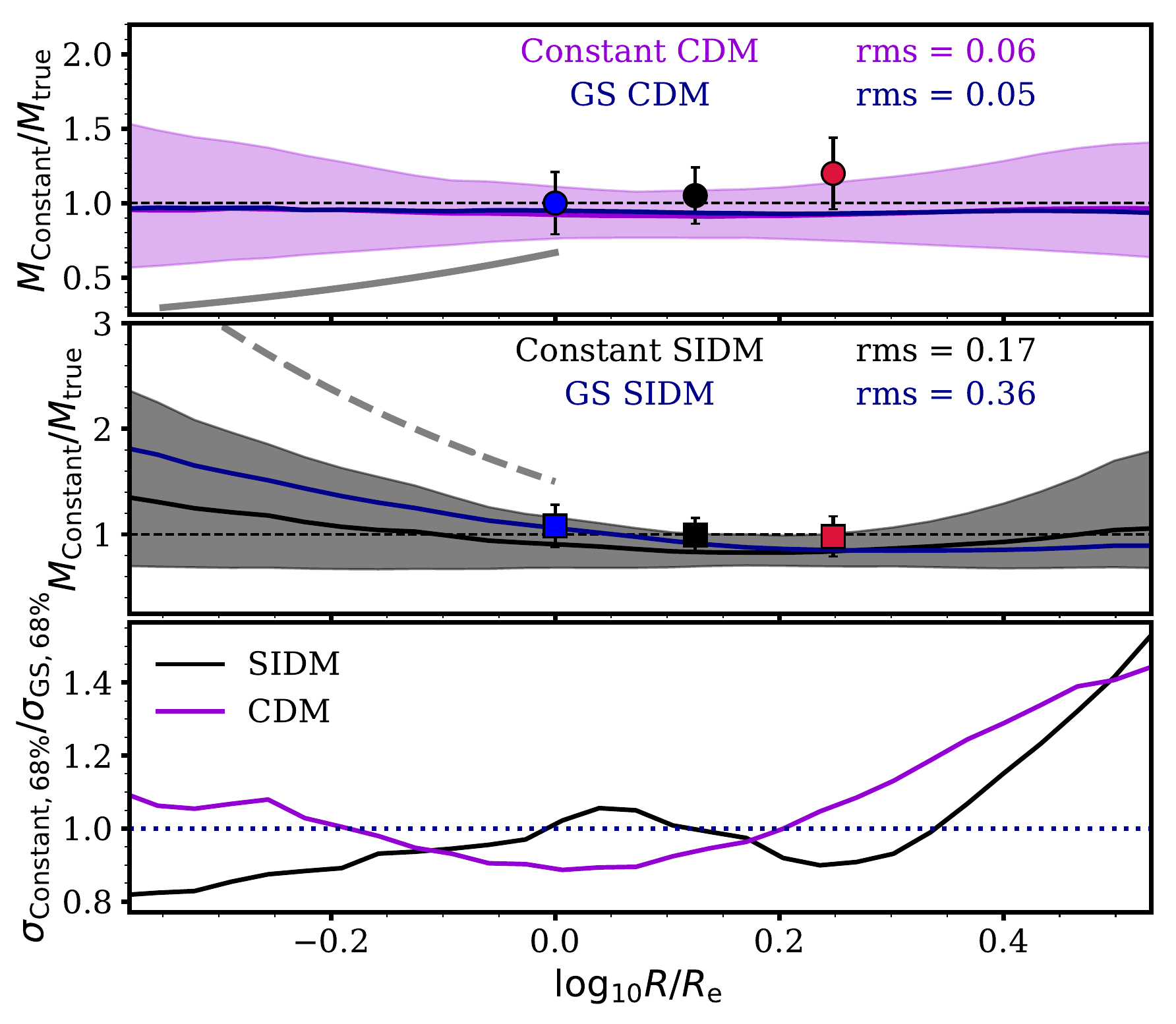}\par
    \includegraphics[width=0.95\columnwidth]{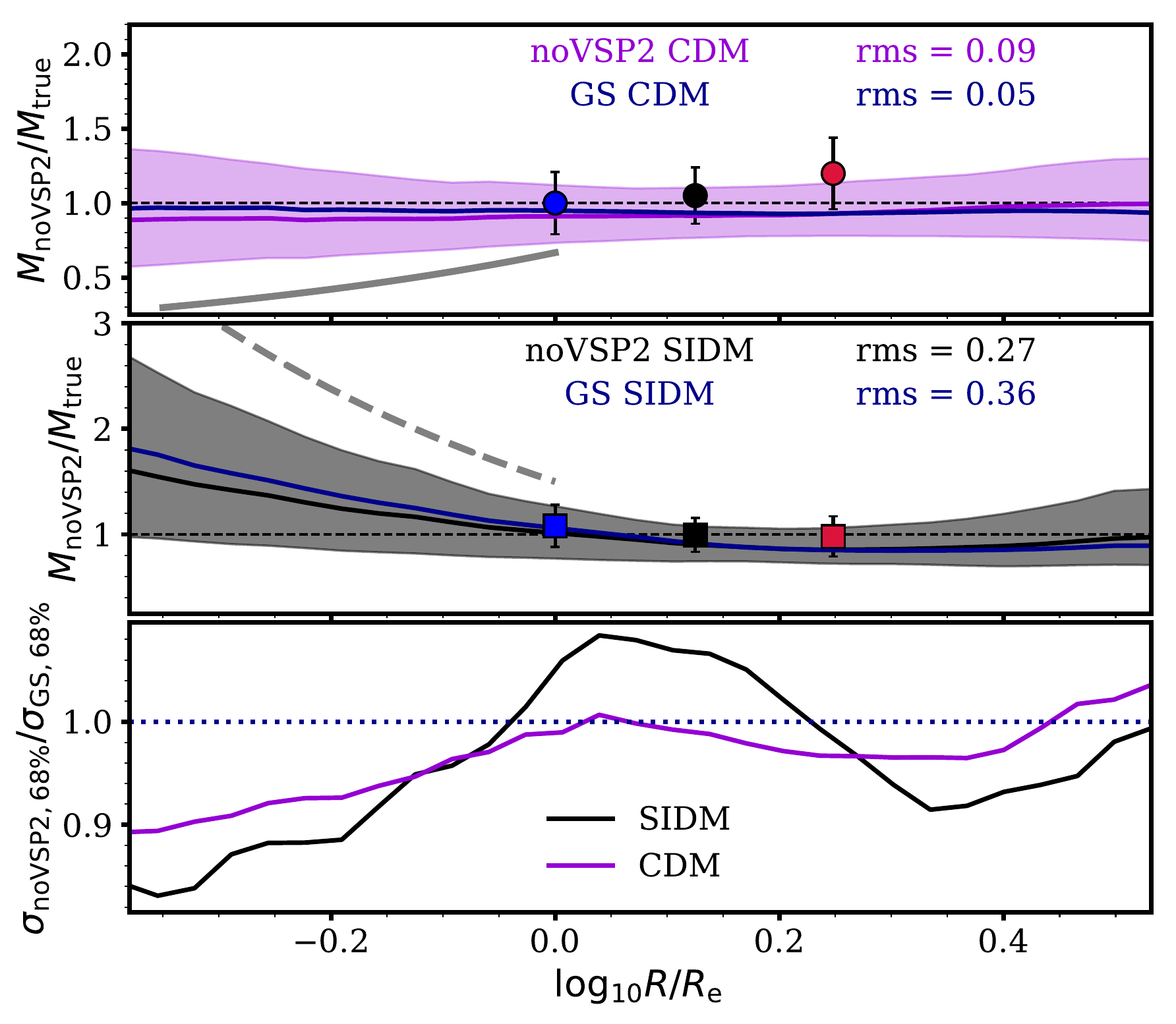}
\end{multicols}

\caption{\textit{Top left:} bias in enclosed mass profiles found assuming a \citet{zhao} profile and using priors as outlined in Table~\ref{table2}, together with post-processing cuts ($\gamma<1$ and $r_s > R_e$). The median for the original {\sc GravSphere} run is shown in blue. The medians for the CDM and SIDM samples are shown with purple (top subplot) and black (middle subplot) lines, respectively. The shaded bands of the same colour show the 68~per~cent confidence limits on the bias, M/M$_{\rm true}$, for CDM (top) and SIDM (middle) samples. The bottom subplot shows the magnitude ratio of the 68 per cent confidence intervals between the \citet{zhao} and default {\sc GravSphere} results. The colour symbols are the \citealt{walkerest} (blue), \citealt{wolf} (black) and \citealt{campbell} (red) estimators. The grey dashed line corresponds to the bias required for an incorrect inference of an NFW cusp, when in reality there is a core, and the solid grey line is the bias required for an inference of a core, when in reality there is an NFW cusp. \textit{Top right:} bias found with {\sc GravSphere} assumptions, but excluding the VSPs (noVSP). \textit{Bottom left:} bias found assuming constant anisotropy and no VSPs. \textit{Bottom right:} bias found with the exclusion of the second virial shape parameter.}
 \label{fig4}

\end{figure*}

We now compare the performance of {\sc GravSphere} to that of more conventional Jeans approaches. We will focus on the enclosed mass profiles, rather than densities. Masses are more robust than their differentials, they are simply measured in simulations and are a fundamental property in Jeans analysis.

\subsubsection{Comparison to \citet{zhao} profile}

We compare the performance of {\sc GravSphere} to the method outlined by \citet{clumpy}, who used the profile proposed by \citet{zhao} to parametrize the dark matter distribution:

\begin{equation}
\displaystyle \rho_{\rm dm} = \frac{\rho_0}{\left({\frac{r}{r_s}} \right)^{-\gamma}\left(1 + \left( \frac{r}{r_s}\right)^{\alpha} \right)^\frac{\beta-\gamma}{\alpha}}    
\end{equation}
where $r_s$ and $\rho_0$ are the scale radius and scale density, $\gamma$ is the inner slope, $\beta$ the outer slope and $\alpha$ governs the steepness of transition between $\gamma$ and $\beta$.

The priors for this run, which we refer to as $\mathrm{Zhao}+\gamma,r_s$, are given in Table~\ref{table2}. The method requires two post-processing cuts: one on $\gamma \leq 1$, which was shown to reduce the overall scatter, and another on $r_s \geq R_e$, which weeds out unphysical models from the fit. The results are shown in the upper left panel of Fig.~\ref{fig4}. 

It can be seen that this method underestimates the enclosed mass by $\sim$10 per cent in the inner regions for CDM dwarfs and is unbiased in the outskirts. In SIDM dwarfs, however, the bias in the centre is less severe than with {\sc GravSphere}. In the inner regions, for CDM and SIDM dwarfs, the scatter has not changed significantly, but outside the half-light radius and in the outer parts the width of the scatter increases by over 50 per cent. This is likely due to the priors (Table~\ref{table2}) allowing a much wider range of outer slopes $\beta$ than permitted by the default {\sc GravSphere} priors (Table~\ref{table1}). The exclusion of virial shape parameters likely also plays a role. We explore this below.

\begin{table}
\centering
\caption{\citet{zhao}$+\gamma,r_s$ MCMC priors and the cuts applied in post-processing}
\begin{tabular}{c|c|c}
Parameter & Prior & Constraint \\
\hline \hline
$ \log_{10}\rho_s$/M$_{\odot}$kpc$^{-3}$   & [5, 13] & \\
$\log_{10} r_s$/kpc    &  [-3, 1] & $\log_{10} r_s \geq \log_{10} R_e$ \\
$\alpha$ & [0.5, 3]&\\
$\beta$ & [3, 7]&\\
$\gamma$ & [0, 1.5]& $\gamma \leq 1$
\end{tabular}
\label{table2}
\end{table}

\subsubsection{{\sc GravSphere}, excluding the VSPs} 

What is gained by including the virial shape parameters? We repeat our {\sc GravSphere} run, this time excluding the two VSPs from Equation~\ref{likelihood}. The results are shown in the top right of Fig~\ref{fig4}. It can be seen that the accuracy in the inner regions has suffered from excluding the VSPs in CDM dwarfs. In SIDM, however, the bias is slightly reduced. As we shall see below, this reduction is consistent with excluding only the second virial shape parameter. The bottom panel shows the ratio of the upper and lower errors (84$^{\rm th}$ and 16$^{\rm th}$ percentiles) compared to the default {\sc GravSphere} run (including the VSPs). It is evident that the scatter has increased beyond the half-light radius. This suggests that {\sc GravSphere} runs that exclude the VSPs result in a wider range of allowed models. Evidently, the inclusion of VSPs plays a key role in minimizing the scatter in allowed anisotropy models, particularly in the outer regions. These results mimic those produced with the priors in Table~\ref{table2}, although the increase in scatter in the outer regions is smaller. This suggests that another significant source of this scatter is, in fact, the wider range of allowed slopes $\beta$.

\subsubsection{Constant anisotropy and no VSPs}
We now explore the performance of {\sc GravSphere} under the assumption of constant anisotropy and no VSPs. From Fig.~\ref{fig2}, it is clear that the vast majority of our simulated dwarfs have nearly constant stellar velocity anisotropy profiles. It is therefore possible that forcing the anisotropy profile to be constant with distance may encourage the MCMC algorithm to select better models. The comparison with {\sc GravSphere} is shown in the bottom left of Fig.~\ref{fig4}. Similar accuracy to {\sc GravSphere} is achieved (RMS = 0.06) across the entire radial range. The errors for CDM dwarfs are similar to {\sc GravSphere} in the innermost regions, but outside R$_e$ the errors are larger. This could be partly due to lack of flexibility as compared to the \citet{baes} profile, but this is also overall consistent with the noVSP run, suggesting a lack of constraint in the outer profile due to the exclusion of VSPs.

For SIDM dwarfs the bias has reduced remarkably (now below 20~per~cent in the inner regions). The scatter has also reduced by $\sim25$~per~cent compared to {\sc GravSphere}. This would suggest that anisotropy profiles are not very well recovered with {\sc GravSphere} and are a significant source of bias for SIDM dwarfs. In Fig.~\ref{fig2} we can see that the true anisotropy profiles are generally enclosed within the confidence limits. These limits are, however, quite large and so the $M-\beta$ degeneracy is not fully broken. The breaking of the degeneracy is forced when imposing a constant form for the anisotropy, resulting in the reduction in bias. Note also that the bias associated with a CDM cusp (dashed grey line in the bottom left of Fig.~\ref{fig4}) is now far above the 68~per~cent limits, compared to the case of {\sc GravSphere}.

We note that, based on the sample of galaxies presented in this paper, our simulations suggest that dwarfs may have anisotropy profiles that are well described by a constant value. For this sample, the assumption of constant $\beta$ is sufficient to accurately recover mass profiles in the innermost regions of dwarfs and this does not require the use of the VSPs.

\subsubsection{Removing VSP2}

In their recent work, \citet{kaplinghat} have opted to exclude the second virial shape parameter, which is extremely sensitive to the behaviour of the fourth velocity moment in the outer regions of the dwarfs (due to the $R^3$ weighting), where the velocity distribution is typically poorly constrained. In {\sc GravSphere}, the uncertainty in the $\langle v^4_P \rangle$ profile is encapsulated within the errors, though these errors are indeed very large. In this subsection we investigate the effects of excluding the second virial shape parameter from our analysis.

The results can be seen in the lower right panel of Fig.~\ref{fig4}. We can see that, for CDM dwarfs, the mass profiles are now underestimated in the inner regions by nearly 10~per~cent, while the scatter is reduced. For SIDM dwarfs, the accuracy has marginally improved and the scatter has also reduced. We conclude that the inclusion of VSP2 in Jeans analysis tends to slightly increase the scatter (due to its large errors), but the accuracy of the mass profiles is improved at the expense of this scatter. 

\subsection{Bias towards cusps in SIDM haloes}
\label{sidmsec}

We now return to the issue of the overestimation of the enclosed mass profiles in SIDM dwarfs. We consider the effects of possible offsets between the centre of mass of the stars and the dark matter as well as the effect of our priors on the recovered mass profiles. 

\subsubsection{Galaxy -- halo offsets in SIDM}

For a profile with a core it is particularly difficult to establish the location of the density centre. If the centre is offset from the true dynamical centre this may introduce a bias in the recovered mass profile. Although it is unlikely that this bias would always cause an overestimation of the true mass, we have investigated the differences between the centre of mass of the stars and the dark matter for each dwarf in our sample. We found that offsets are present, with a typical magnitude of $80\pm35$~pc for the CDM sample. For SIDM halos the offsets are particularly extreme, up to 1~kpc in size. We repeated our analysis, comparing the mass profiles recovered by {\sc GravSphere} to those computed directly from the simulation, now centering the `true' profiles at the centre-of-mass of the dark matter. We found no discernible differences from the results presented in Fig.~\ref{fig3}. Centering on the centre of potential instead also did not change our results. We conclude that galaxy--halo offsets have no significant effect on the results presented in this work. This is because these offsets are rather small compared to the half-light radii of SIDM dwarfs.

\subsubsection{Using priors that favour a core}

\begin{figure}
        
		\includegraphics[width=\columnwidth]{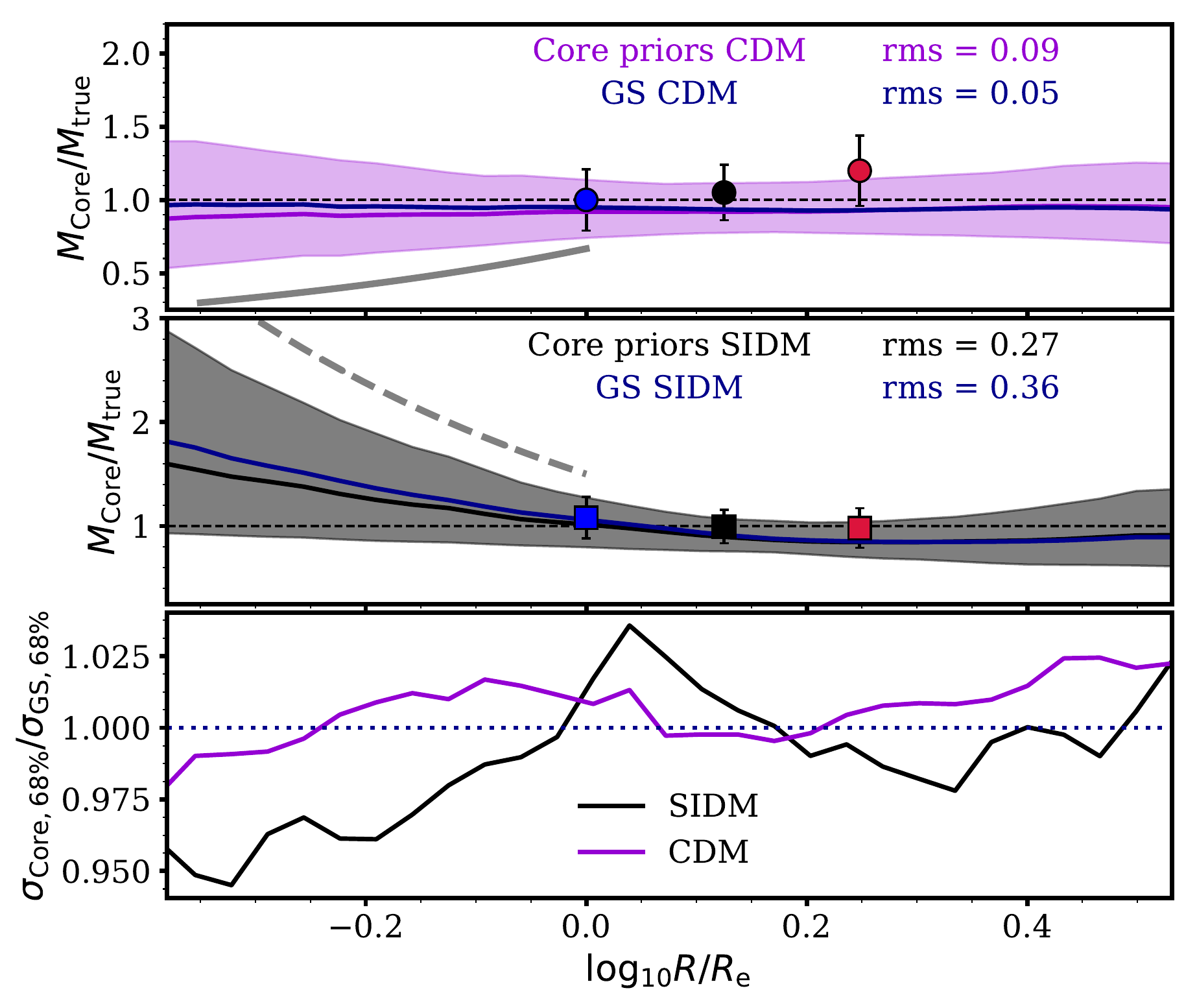}\par

		\caption{Bias found using wider priors on the slopes $\gamma_j$ that allow a `hole' in the central regions. All values of $\gamma_j<0$ output by the MCMC are then equated to 0. The coloured symbols and grey lines are as in Fig.~\ref{fig4}.}

		\label{fig5}
	\end{figure}

The systematic overestimation of enclosed mass at small radii has been encountered previously in the work of \citet{readdraco}. The priors in Table~1 may bias the results towards more cusp-like values when the data are not sufficiently constraining, as a core (with $\gamma = 0$) lies on the boundary of the allowed range of slopes (see Fig.~\ref{AppFig1}). In \citet{readdraco}, a different set of priors was introduced, which extends the parameter space of the dark matter density slopes $\gamma_j$ such that {\sc emcee} walkers are more likely to probe regions of space that are compatible with a core. This is achieved by allowing the slopes to range between $-2 <\gamma_j < 3$. Note that $\gamma_j < 0$ corresponds to a `hole' in the central regions of the dwarf. While not implausible in principle, in this work we assume that, in these extreme cases, the dark matter distribution has a core. When computing the confidence limits of the enclosed mass and density distributions in each dwarf, we thus fix the $\gamma_j$ position of any walkers venturing into the space where $\gamma_j<0$ to $\gamma_j = 0$.

The results of this may be seen in Fig.~\ref{fig5}. For SIDM dwarfs, we can see that the bias along the entire radial range has been reduced to $\sim25$~per~cent and the scatter is reduced by $\sim10$~per~cent in the inner regions. We also display the results for individual SIDM dwarfs in Fig~\ref{AppFig3}, where a reduction in bias compared to Fig.~\ref{fig2b} is evident. For CDM dwarfs, the new priors slightly bias the recovered mass profiles towards cores, although the true profiles still lie within the 68~per~cent confidence levels. This is consistent with the findings of \citet{readdraco}.

We conclude that the lack of exploration of the models with $\gamma_j = 0$  by {\sc GravSphere} is clearly an issue, however the data still lack sufficient constraining power to prefer cores over cusps in SIDM.

\subsection{Using all available stars}

\begin{figure}
        
		\includegraphics[width=\columnwidth]{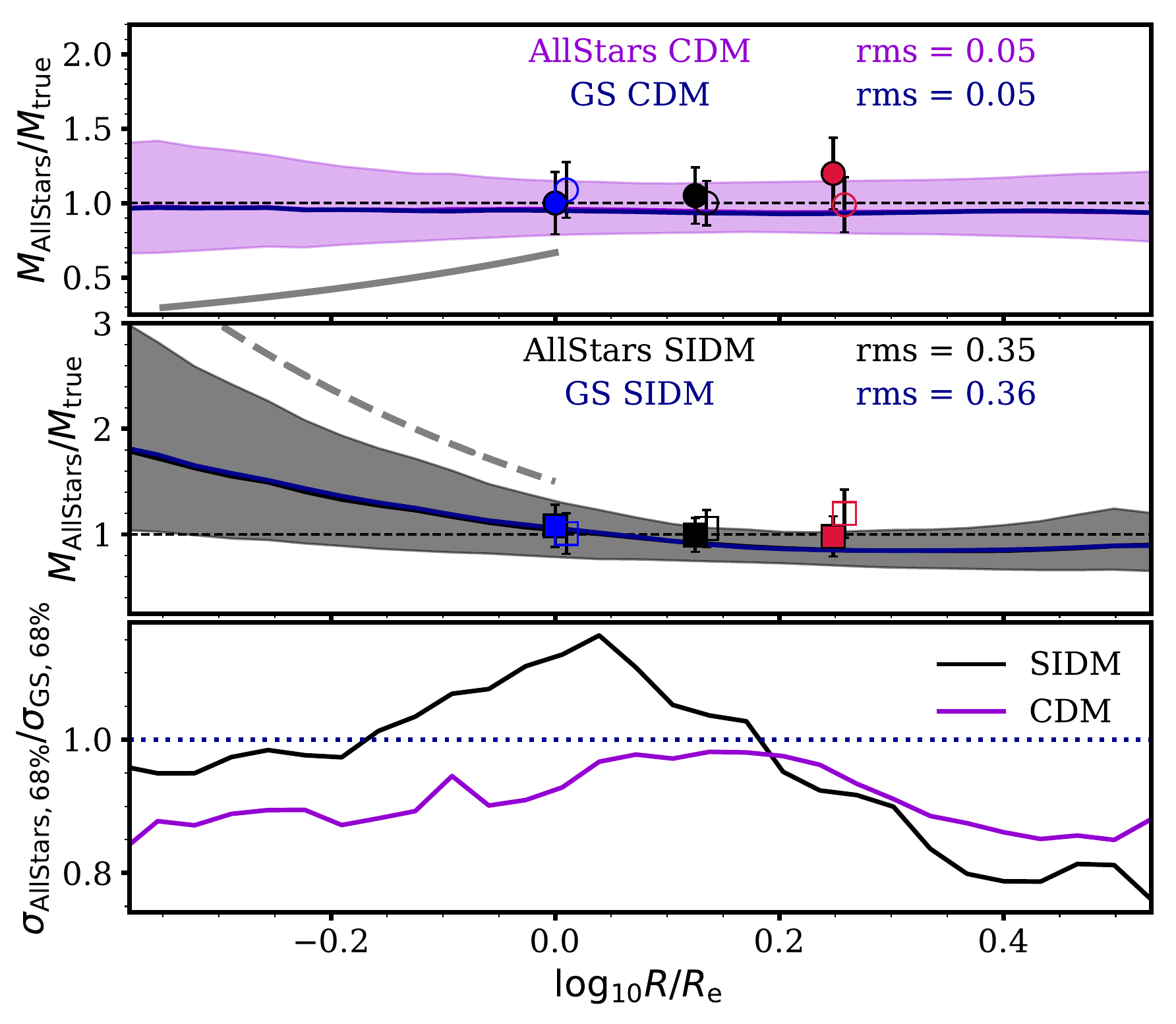}\par

		\caption{Bias found under standard {\sc GravSphere} assumptions, but using all available stars in each dwarf. The symbols are as in Fig.~\ref{fig4}. The additional empty symbols representing the accuracy of mass estimators were evaluated using all available stars within $2R_e$.}

		\label{fig6}
	\end{figure}
In this section we briefly explore how much information is gained by including all stars within $2R_e$ of each dwarf (ranging from $10^3$ to $10^4$ stars for CDM), as opposed to a Fornax-like sample of $500 - 2500$~stars. We note, however, that for the majority of SIDM dwarfs the available samples are less than 1000 stellar particles. 

The results of this can be seen in Fig.~\ref{fig6}, where it is clear that the accuracy in the inner regions has slightly improved and the size of the errors is reduced by $\sim 10$~per~cent at smaller radii and closer to $\sim20$~per~cent at large radii. Larger samples of data from future spectroscopic surveys will undoubtedly reduce the uncertainties associated with recovered dark matter mass profiles; however this improvement is expected to be rather small when using line-of-sight data only. Moreover, {\sc GravSphere} is able to achieve the same level of bias with present data samples. Further reduction in uncertainty is more likely to come from exploiting kinematically distinct stellar populations or proper motion data \citep{gravsphere}.

\subsection{Sources of bias and scatter in {\sc GravSphere}}

We now explore possible origins of scatter in Fig.~\ref{fig3}. Projection effects, asphericity and tides are of particular interest \citep{genina}. 

\begin{figure}
        
		\includegraphics[width=\columnwidth]{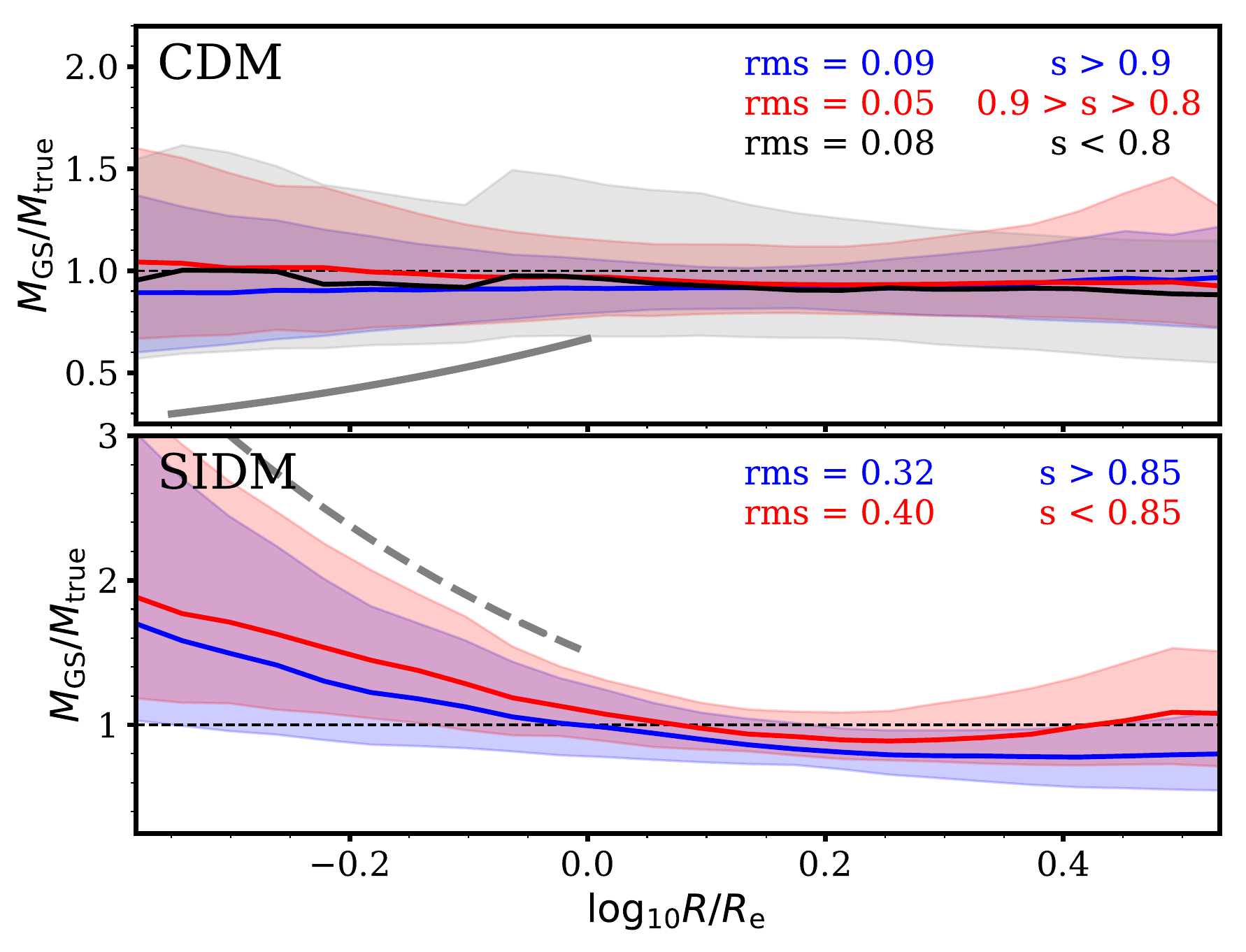}\par
		\includegraphics[width=\columnwidth]{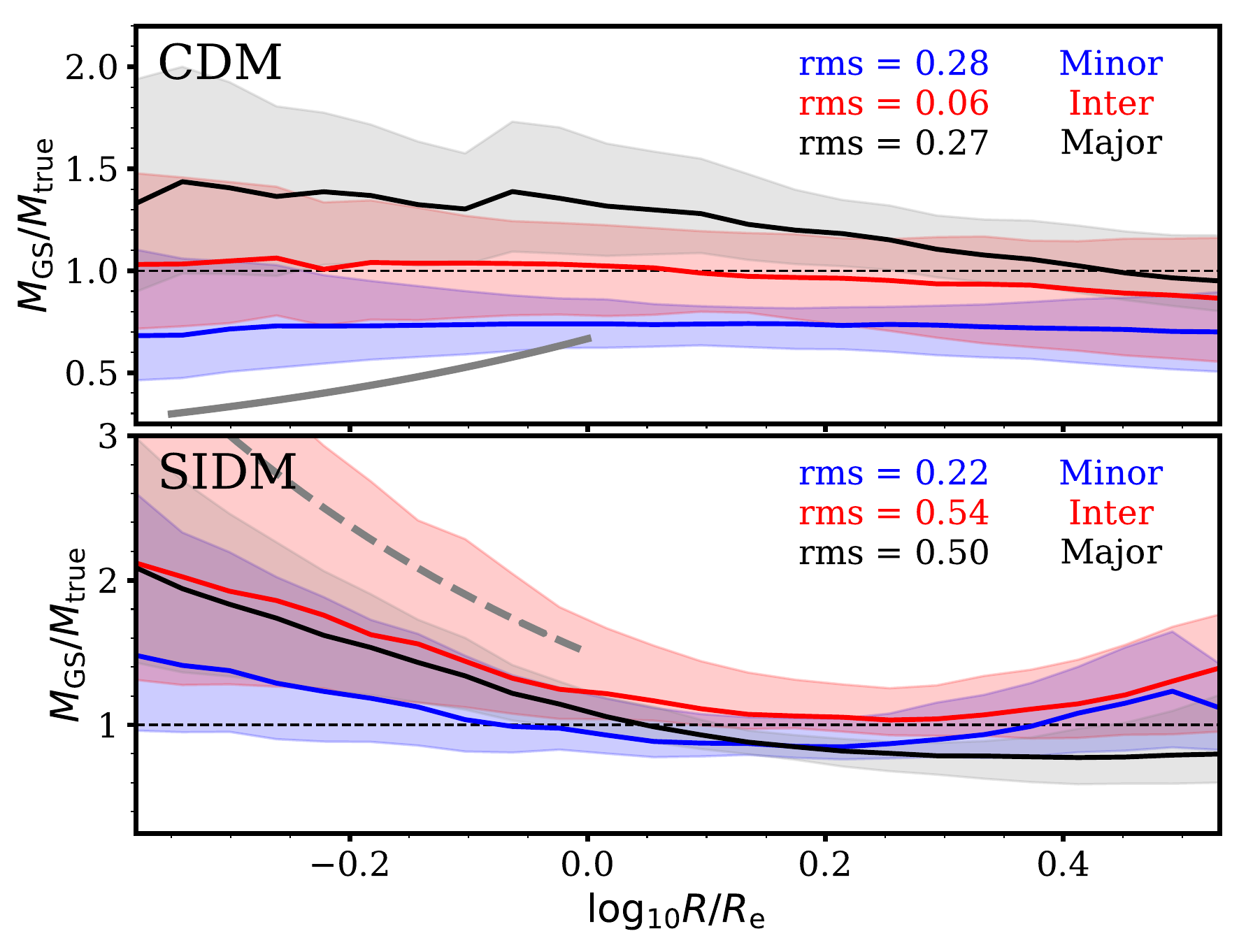}\par

		\caption{\textit {Top:} bias in mass profiles returned by {\sc GravSphere} for dwarfs grouped by their minor-to-major axis ratio $c/a$. The top panel shows the CDM sample and the bottom panel shows the SIDM sample. \textit{Bottom:} line-of-sight effects on the accuracy of the mass profile recovery by {\sc GravSphere}. The sample of galaxies is split into those viewed along the minor (blue line and bands), intermediate (red) and major (black) axis. Only dwarfs with $s < 0.8$ for CDM and $s < 0.85$ for SIDM are shown. The dotted black line in each plot represents unbiased results. The solid grey line corresponds to the bias expected in order to incorrectly infer a core within the half-light radius of a cuspy dark matter halo. The dashed grey line corresponds to the bias expected to incorrectly infer a NFW cusp, when in reality there is a core.}

		\label{fig7}
	\end{figure}

\begin{figure*}
        \centering
		\includegraphics[width=2\columnwidth]{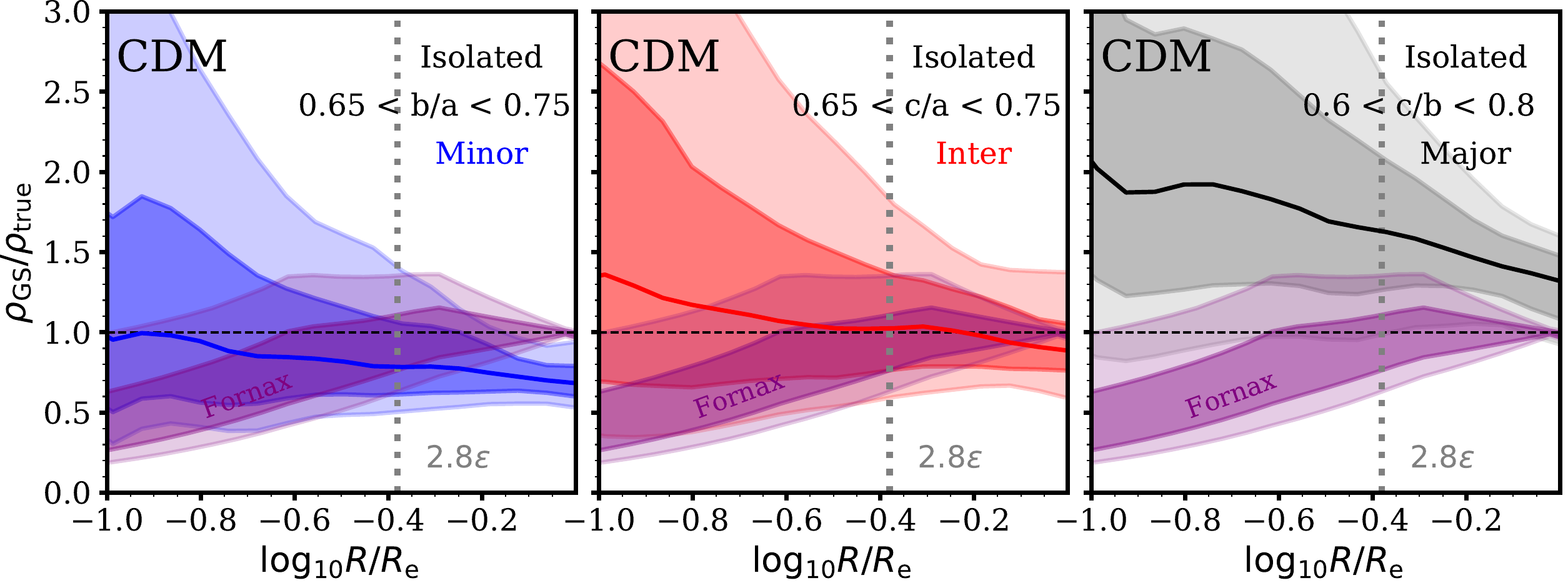}\par

		\caption{The bias in density profiles inferred by {\sc GravSphere} for 24 isolated CDM dwarfs that match the criteria of having an on-the-sky ellipticity $e\sim0.3$ when viewed along their minor (left, blue), intermediate (middle, red) and major (right, black) axes. The dark and light bands show the 68 and 95~per~cent confidence levels, respectively. The dotted grey line shows the average value of $2.8\epsilon/R_e$. The purple line and bands shows the inferred density profile for Fornax from \citet{heating}, divided by a cuspy profile, $\rho \propto r^{-1}$, which is normalized to the recovered density of Fornax at $R_e$.   }

		\label{fig8}
	\end{figure*}

\subsubsection{Line-of-sight effects}
\label{loseffects}

How does asphericity affect the accuracy of the enclosed mass profiles recovered by {\sc GravSphere}? In the top panel of Fig.~\ref{fig7} we split our sample of dwarfs in bins of minor-to-major axis ratio $s = c/a$. It can be seen that, on average, dwarfs of all asphericities in CDM have mass profiles recovered to better than 10~per~cent; however it is also clear that the scatter is much larger for more aspherical objects ($s < 0.9$). Our sample has been chosen to contain galaxies projected along their three principal axes. We can therefore split our sample into 3 categories: dwarfs seen along the minor, major and intermediate axes, and examine the accuracy of {\sc GravSphere} in recovering mass profiles in each case. For this, we select dwarfs with $s<0.8$ in CDM (6 galaxies) and $s < 0.85$ in SIDM (3 galaxies) \footnote{We have considered projection effects for the more spherical dwarfs in the sample and found that they are irrelevant in those cases.}. 

The results are shown in the bottom panel of Fig.~\ref{fig7}. For CDM dwarfs, there is a clear distinction between mass profiles obtained when viewing along the minor (blue), intermediate (red) and major (black) axis. Along the intermediate axis the mass profiles are generally unbiased, but they are underestimated along the minor axis and overestimated along the major axis. On average, this over- and under-estimation is of magnitude $\sim30$~per~cent. For prolate systems, this is consistent with the variation of line-of-sight velocity dispersion when viewing along the 3 principal axes.

The picture is somewhat different for SIDM dwarfs, where the mass profiles are overestimated more significantly when viewed along the intermediate axes. We note that this sample contains only 3 dwarfs and their 3 projections. We found that two of these have dark substructure present within their halos, resulting in inflated velocity dispersion along the intermediate axis of these dwarfs.

We have also considered the effects of dark matter halo asphericity. We found that the shapes of dark matter haloes are typically consistent with the stellar component, apart from a number of dwarfs with aspherical stellar distributions, for which the dark matter component was generally \textit{less aspherical} than the stars. We conclude that the systematics are driven by the asphericity of the stellar component.

\subsubsection{The core-cusp problem in Fornax}

Results presented in \citet{heating} suggest that Fornax may have an inner core, based on the low inferred dark matter density and the inner slope, $\gamma_0 =0.3^{+0.28}_{-0.21}$, below 0.25$R_e$ (see also \citealt{goerdt}, \citealt{cole_glob_clust} and \citealt{pascale19}). Earlier, we have established that, for aspherical stellar distributions, line-of-sight effects may bias the recovered mass and density profiles. What do these results mean for the Fornax dwarf galaxy?

 Fornax has a measured on-the-sky ellipticity $e\simeq0.3$, matched by only a few satellites in our sample. We therefore focus instead on a sample of 24 isolated, dispersion-supported dwarfs from the APOSTLE simulations in CDM. For consistency with Fornax, these dwarfs were chosen to have an ellipticity $e\sim0.3$ when viewed along at least one of their principal axes. Otherwise, we marginalize over various shapes of the stellar and dark matter components. The axis ratios for these galaxies can be seen on the right panel of Fig.~\ref{fig1} (grey triangles).
 
 Fig.~\ref{fig8} shows the bias in density profiles recovered with {\sc GravSphere} for dwarfs viewed along the minor, intermediate and major axes (and $e\sim0.3$ in each case). The results are shown for radii between $0.1R_e$ and $R_e$, a key region where core formation would be apparent. Note that we have now gone below 2.8$\epsilon$ (grey dotted line), close to the softening length of our simulations. We must be wary of the discreteness noise contribution to the scatter at these radii. The purple bands show the 68 and 95~per~cent confidence levels of the Fornax density profile recovered with {\sc GravSphere} \citep{heating}, divided by the profile in Equation~\ref{powerlawprof}, with $\rho_0$ equal to the density recovered by {\sc GravSphere} for Fornax at $R_e$.
 
 It is important to point out that, due to the mass resolution in APOSTLE, the innermost regions of our simulated dwarfs are insufficiently sampled, compared to the real Fornax data, which results in systematically larger errors in the recovered density profiles. Indeed, the typical span of the 68~per~cent confidence intervals below $0.25R_e$ in Fig.~\ref{fig2} is 0.5-0.6 in $\log_{10}\rho$, while for Fornax it is below 0.4, with the size of the errors being approximately the same for simulations and Fornax above 2.8$\epsilon$. In order to compensate partially for this, in Fig.~\ref{fig8} we scale the errors for each dwarf by the ratio between the fractional errors (the span of the 68~per~cent confidence intervals divided by the median) in a given APOSTLE dwarf and Fornax. This scaling, however, is not able to alleviate the large growth in uncertainty below $R = 2.8\epsilon$ for the mock data. Tackling this will require higher resolution simulations that are beyond the scope of this work.
 
A particularly striking result in Fig.~\ref{fig8} is that, above 0.1$R_e$, the inferred density distribution in Fornax appears to be consistent within its 95~per~cent confidence intervals with a cuspy dark matter halo viewed along any of the three principal axes. However, \citet{heating} disfavour this interpretation -- at least in the context of a $\Lambda{\rm CDM}$ cosmology. This is because it requires an uncomfortably low pre-infall halo mass\footnote{$M_{200}$ is the enclosed mass at $r_{200}$, the radius below which the mean density is 200 times the critical density of the Universe.} for Fornax ($M_{200}\simeq 5 \times 10^9$\,M$_\odot$) as compared to expectations from abundance matching ($M_{200}\simeq 2 \times 10^{10}$\,M$_\odot$; \citealt{readerkal}). It also requires a smaller halo concentration than is characteristic of such a halo mass in $\Lambda{\rm CDM}$ \citep{mass-conc}. It remains to be seen whether a cuspy Fornax in a low pre-infall mass and concentration halo can self-consistently explain the relatively low density of Fornax in the context of $\Lambda{\rm CDM}$ cosmology. We will revisit this in future work, but note it here as a potential caveat to the conclusion of \citet{heating} that favours a core in Fornax in a more massive pre-infall halo.

Finally, Fig.~\ref{fig8} shows that the bias in {\sc GravSphere} tends to remain fairly constant below the half-light radius. This is an encouraging result, since, in the context of the core-cusp problem, we must be particularly wary of any radially-dependent systematics. If higher resolution simulations find similarly little radial dependence in the systematic errors down to 0.25$R_e$ and below, then this would suggest that the dominant source of uncertainty, at present, in estimating the inner dark matter density profile of the Fornax dwarf lies in the random sampling error (i.e. the number of stars with kinematic data), rather than the systematic error due to asphericity and the projection of Fornax on the sky.

\subsubsection{The effect of tides}
\label{tidessection}

\begin{figure}
        \centering
		\includegraphics[width=\columnwidth]{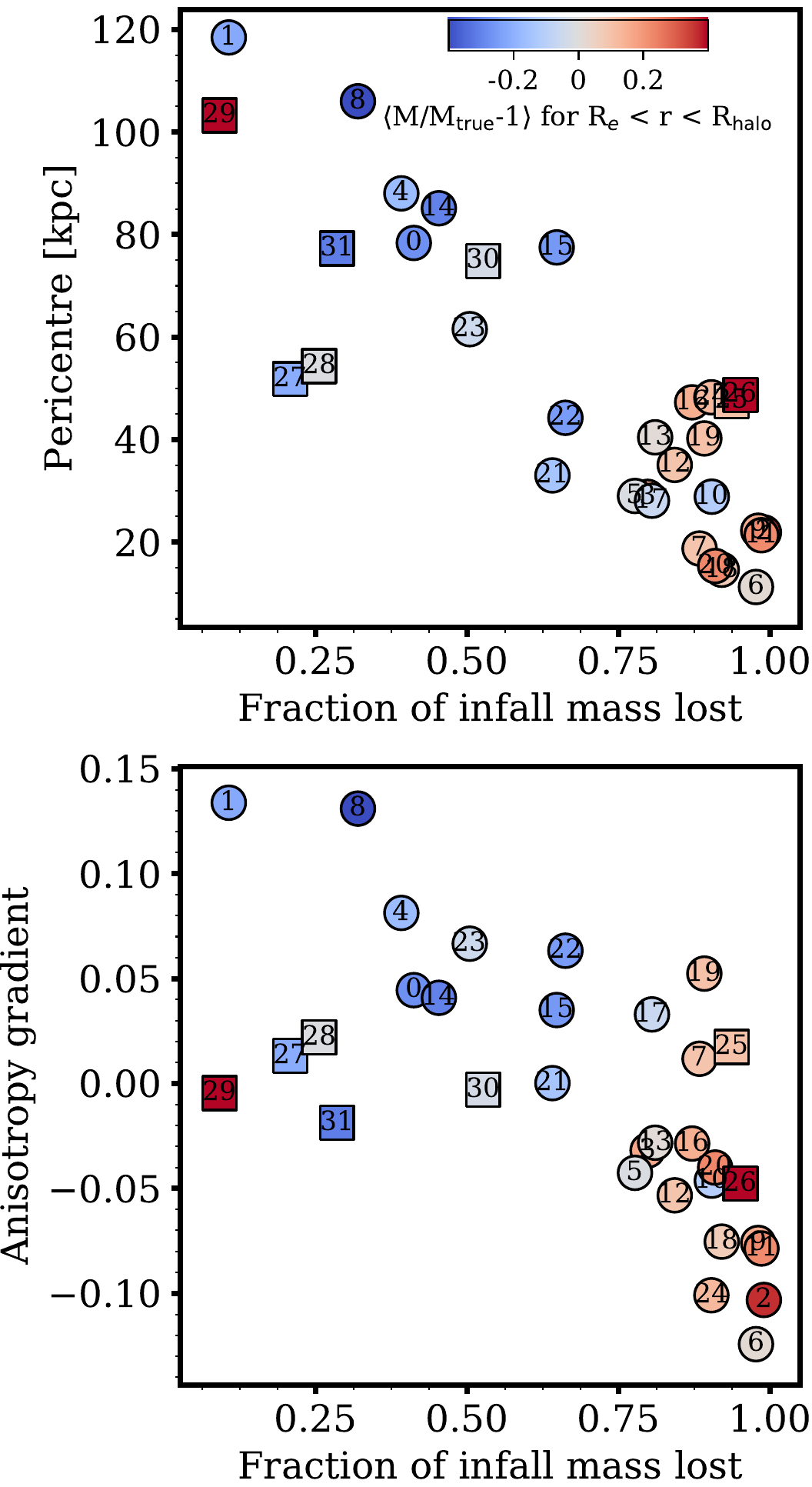}
		\caption{\textit{Top:} the pericentre of the dwarf galaxy orbit as a function of the fraction of dark matter mass that has been lost since infall. The points are coloured by the mean bias of the recovered mass profile above the half-light radius and below the spatial extent of the dark matter halo. This is averaged for the three principal axes. Red indicates overestimation and blue indicates underestimation.  Circles represent CDM dwarfs and squares the SIDM dwarfs. \textit{Bottom:} the radial change in anisotropy, quantified by an anisotropy gradient measured below and above the half-light radius, as a function of the fraction of dark matter mass lost through stripping. }

		\label{fig9}
	\end{figure}

Dwarfs in our sample were selected to be satellites and these are susceptible to tidal stripping by their host galaxy. In this section we explore whether the mass profiles recovered by {\sc GravSphere} could be affected by tides. 

In the top panel of Fig.~\ref{fig9} we plot the pericentre of the satellite orbits as a function of the dark matter mass lost since infall. We find the pericentres by interpolating the position of each dwarf with respect to its host with a cubic spline. This method may underestimate the pericentres (see \citealt{richings}); however, for all of our dwarfs we see little difference between pericentres found through the linear and cubic splines. Moreover, the majority of the dwarfs have infall times of $\sim$8~Gyr, such that typically 2-3 orbital periods are available for pericentre calculation, with snapshots having shorter temporal spacings at early times. We define the infall time as the snapshot at which the subhalo has its maximum dark matter mass. As expected, dwarfs with smaller pericentres tend to lose larger fractions of their mass. 

In the bottom panel of Fig.~\ref{fig9} we show anisotropy gradients as a function of lost dark matter mass. We measure the anisotropy gradients between two points: the mean stellar particle radii below and above the projected half-light radius. A clear trend is evident, whereby tides cause more tangential anisotropy in the outer parts of galaxies. This is due to the preferential stripping of the stars moving on radial orbits \citep{henon,keenan,radialstrip, readtides,donghia}. We note the apparent simplicity of the two relations in Fig.~\ref{fig9}, excluding perhaps the SIDM dwarfs, the orbits of which are more isotropic and the anisotropy is approximately constant.

We colour the points in Fig.~\ref{fig9} by the mean bias, $\langle M/M_{\rm true}-1\rangle$, measured for all radii outside $R_e$ and below the spatial extent of the halo. This is averaged for the three lines of sight along the principal axes of the dwarfs. We note that the bias is dominated by the line of sight closest to that which points from the centre of the host galaxy to the dwarf. This is indeed where we expect to see the largest effects on the mass modelling due to the ongoing process of tidal stripping \citep{unboundmodelling}. 

It can be seen in Fig.~\ref{fig9} that for dwarfs that have undergone stronger tidal effects the mass profile is overestimated. This is unsurprising given that tides will result in the steepening of the outer slope \citep{penarrubiatides}, beyond the $0 < \gamma_{j} < 3$ permitted by our priors. Note for example, the increased accuracy in the outer mass profile when using the \citet{zhao} dark matter parametrization (top left of Fig.~\ref{fig4}), where such steep slopes are allowed. An exception from this trend is SIDM Galaxy~29, where the mass profile is overestimated in the outer parts, yet tidal effects seem to be less significant. Nonetheless, the steepening of the density profile of this dwarf beyond $\gamma=3$ is evident in Fig.~\ref{fig2}.

We note the average underestimation of the outer mass profiles in dwarfs with weaker tidal effects. It is unclear how significant this result is, given that a number of these dwarfs (Galaxies 0,1,8,23) are amongst the most aspherical objects in our sample and thus are subject to other sources of bias. 

Most importantly, we point out that we found no significant trend for the accuracy of the enclosed mass profile below $R_e$ in our sample of dwarfs with tidal effects. This suggests that Jeans analysis is a valid method of mass modelling for dwarfs susceptible to tidal interactions, provided the impact of stars that are in the process of being tidally stripped is minimized. Here we achieve this by limiting our sample of stellar particles to those within $2R_{2D}$, as well as only using the stellar particles considered as `bound' by the subhalo finder. The study of the performance of {\sc GravSphere} on a realistically contaminated sample of stars is certainly warranted and has been addressed, with a smaller sample of galaxies, by \citet{gravsphere}.

\subsection{Identifying failing models}
	
Is it possible to identify the cases where {\sc GravSphere} produces a biased result? In \citet{gravsphere}, it was shown that for triaxial haloes it is possible to tell that the model has been unsuccessful through the value of the $\chi^2$. Is this true for our sample of dwarfs?

In Fig.~\ref{fig10} we display the mean number of standard deviations to the true mass profile from the profile recovered by {\sc GravSphere}, computed within the half-light radius of each dwarf, as a function of the total normalized $\chi^2$ value (i.e. we divide the $\chi^2$ for the surface density profile and the line-of-sight velocity dispersion profile by the total number of photometric and kinematic bins, respectively). The points are coloured by the number of standard deviations from the `truth', with red indicating an overestimate and blue indicating an underestimate. The grey shaded regions, highlighting the $1\sigma$ and $2\sigma$ intervals form the true mass profile, are labelled by the fraction of our galaxies falling within there regions. We thus confirm that in just under 60~per~cent of the dwarfs in our sample, {\sc GravSphere} returns the true mass profile within $1\sigma$ and in just under 90~per~cent within $2\sigma$, with the worst results encountered for SIDM dwarfs and the most aspherical objects in our sample.

Strictly speaking, we would expect to classify models with $\chi^2>4$ as poor fits. Indeed, we see that for the values of $\chi^2\gtrsim3.5$ (black dashed line) a stronger bias is observed. In general, there is a significant scatter the quality of results for dwarfs with $\chi^2\gtrsim3.5$. Among these galaxies we see the particularly aspherical dwarfs and some SIDM galaxies (which are biased towards cusps), but not all. Evidently, this cut in $\chi^2$ may be used for the selection of models that are more likely to be unbiased, such that the scatter in the accuracy of mass profiles returned by {\sc GravSphere} is minimized. 

\section{Conclusions}
\label{conclusions}

	\begin{figure}
        \centering
		\includegraphics[width=\columnwidth]{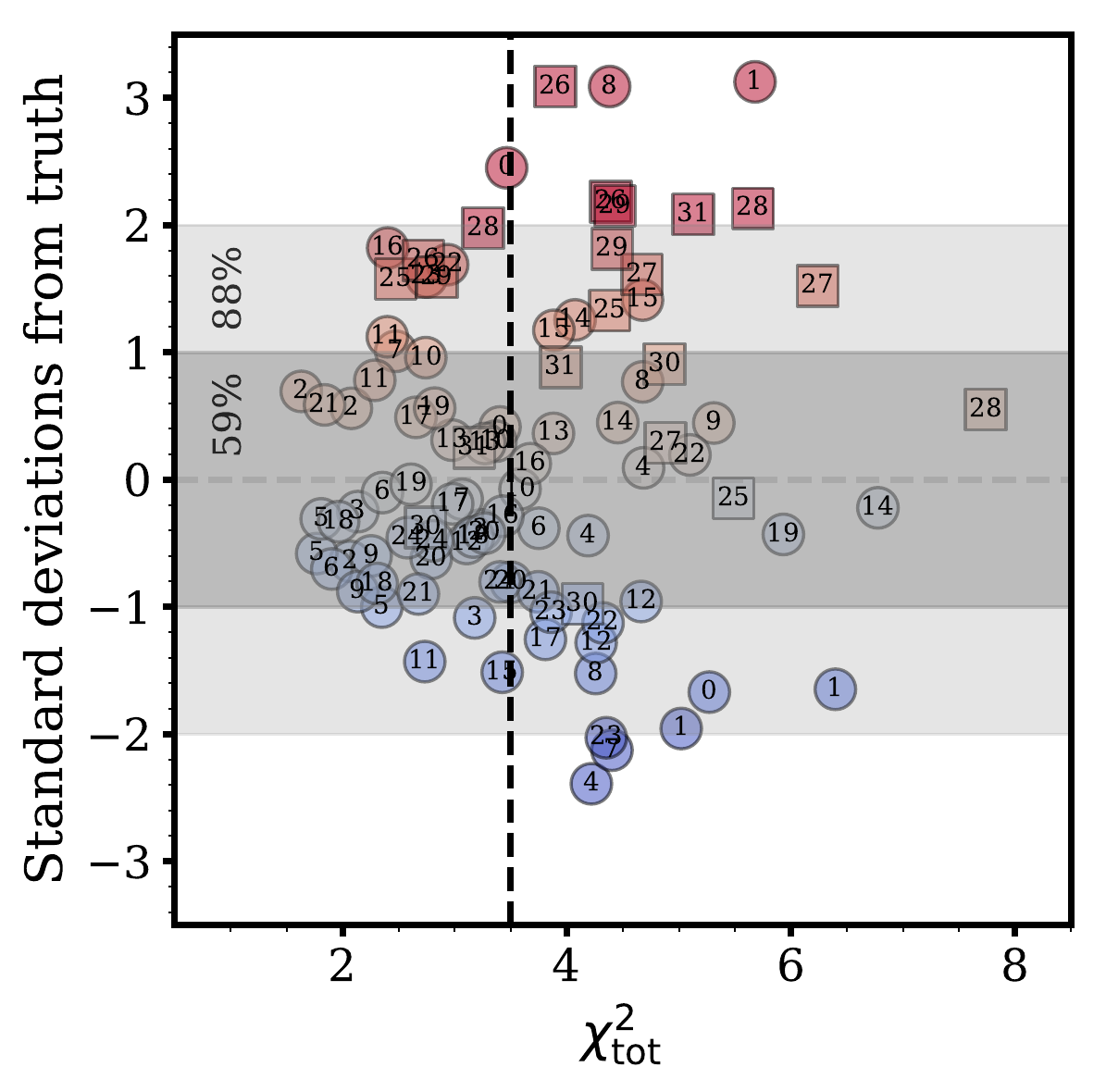}

		\caption{Number of standard deviations between the {\sc GravSphere} result and the true mass profile, computed below the half light radius and above 2.8$\epsilon$, as a function of total, reduced $\chi^2$. CDM dwarfs are shown with circles and SIDM dwarfs with squares. Red highlights an overestimation of true mass and blue an underestimation ($y$-axis values). The grey dashed line indicates unbiased results. The black dashed line shows our suggestion for a `cut' in $\chi^2$ to separate `successful' and likely biased models. The shaded regions display the 1$\sigma$ and 2$\sigma$ regions and are labelled by the percentage of our sample, where the true mass profile is contained within these regions.}
		\label{fig10}
    \end{figure}

Dwarf spheroidals are some of the best objects in which to study dark matter due to their proximity and high mass-to-light ratios. With the increasing availability of high-quality spectroscopic, photometric and proper motion data, studies of mass modelling methods and their limitations using realistic $N$-body simulations are certainly becoming more important in our efforts to narrow down the identity of dark matter and its behaviour on small scales. Here we presented such a study for {\sc GravSphere}, a higher-order non-parametric Jeans analysis method \citep{gravsphere}.

First, we selected a sample of 32 dwarf galaxies from a suite of cosmological hydrodynamic simulations in $\Lambda$CDM and SIDM cosmologies. These simulated galaxies were chosen to resemble classical Local Group dwarfs like Fornax. We then applied {\sc GravSphere}, with its standard set of priors, to each of these dwarfs. We present the following findings:
\begin{itemize}
    \item Within the key region inside the projected half-light radius, where dark matter cores form in some simulations \citep{navarroeke,nihaocores,firecores,firecores2,alltheway,pontzen,alejandrocores}, the enclosed mass distributions are recovered within the 68 per cent confidence limits for $\sim$60~per~cent of the dwarfs in our sample and within the 95~per~cent confidence limits for $\sim90$~per~cent of the dwarfs.

\item For our sample of CDM dwarfs, {\sc GravSphere} returns unbiased mass profiles (RMS = 0.05) along the radial range of (0.4-3)${\rm R_e}$, but with $\sim$ 50 per cent scatter in the innermost regions and $\sim$ 25 per cent scatter at the projected and deprojected half-light radii; this is comparable to standard mass estimators (see the left panel of Fig.~\ref{fig3}). In comparison to other Jeans methods, {\sc GravSphere} achieves a more consistent performance across the radial range considered and, typically, has smaller scatter in the recovered mass profiles (see Fig.~\ref{fig4}).

\item The density profiles for our CDM sample recovered by {\sc GravSphere} are also accurate to better than 10~per~cent (RMS = 0.07) and exhibit a scatter of 30 per cent (see the right panel of Fig.~\ref{fig3}). Within the 68~per~cent confidence levels this is sufficient to reject cores that form on the scale of the half-light radius (when, in reality, there is a cusp). Due to the spatial and mass resolution of our simulations we were only able to test {\sc GravSphere} in regions outside 380~pc; however, if the uncertainty in the density profile does not increase for regions near 100~pc, it should be possible to separate core and cusp-like densities for Fornax-like dwarfs that have undergone complete core formation on the scale of the half-light radius, as described in \citet{alltheway}, provided the central density has not been reduced by tides and the pre-infall halo mass is well constrained. 

\item For the sample of SIDM dwarfs with interaction cross-section of $\langle\sigma/m\rangle=10$~$\rm{cm^2g^{-1}}$, which have cores on the scale of the half-light radius, we find that {\sc GravSphere} is biased towards cuspy models. The density is overestimated by $\sim20$~per~cent in the central regions. We show that this bias is relieved when imposing the correct form of anisotropy ($\beta = \beta_0$), suggesting that the data are not sufficiently constraining to break the $M-\beta$ degeneracy for these dwarfs (see the bottom left panel of Fig.~\ref{fig4}). Moreover, our priors do not allow a full exploration of parameter space when there is a core in the central density distribution. Widening the priors to allow `holes' in central regions leads {\sc GravSphere} to a reduction in the bias for the SIDM sample (see Fig.~\ref{fig5}), although at the expense of slightly biasing the CDM sample towards cores, with a 10~per~cent underestimation in the inner regions. 

\item We explored the benefits gained by having larger stellar samples for our dwarfs (see Fig.~\ref{fig6}). We found that using all available stellar particles within 2R$_e$ mildly improves the accuracy of the mass profiles and reduces the scatter by $\sim15$~per~cent for the CDM sample and $\sim5$~per~cent for the SIDM sample (where we are limited in the number of available stellar particles due to mass resolution). 

\item Our simulations suggest that Fornax-like dwarfs may have anisotropy profiles consistent with a constant value, $\beta = \beta_0$. In fact, for dwarfs in CDM, assuming a constant $\beta$ and no VSPs results in a similar bias and scatter in the inner regions as {\sc GravSphere} (with an increase in the size of the errors in the outer parts). For SIDM dwarfs, the assumption of constant $\beta$ significantly reduces the bias and scatter at small radii compared to {\sc GravSphere}, without the need to change the priors on the dark matter density slopes to allow for more models with cores (see the bottom left panel of Fig.~\ref{fig4}).
\end{itemize}

We have explored the reasons for the scatter in {\sc GravSphere}'s performance. We found the following:
\begin{itemize}

\item The scatter in the accuracy of the recovered mass profiles is largest for objects that are particularly aspherical (see top panel of Fig.~\ref{fig7}). In our CDM sample of dwarfs with the minor-to-major axis ratio, $c/a\sim0.7$, galaxies viewed along their intermediate axis typically have their masses recovered accurately (RMS = 0.05) compared to objects viewed along the minor axis (where the mass is underestimated by $\sim30$~per~cent) or the major axis (where the mass is typically overestimated by $\sim$30~per~cent). If Fornax has a sphericity of $c/a\simeq0.7$ and is viewed along the intermediate axis, {\sc GravSphere} is expected to accurately recover the mass (and the density) profile out to the half-light radius. 

\item We have explored the effect of tides on the performance of {\sc GravSphere}. We found no significant effect on the recovery of the mass profiles below the half-light radius. However, we did find that the mass profiles of systems which are more significantly affected by tides are typically overestimated in the outer regions (see Fig.~\ref{fig9}), primarily due to the imposed priors on the outer density slope. This suggests that Jeans analysis is still valid for systems affected by tidal interactions, provided the impact of stars in the outer regions, which could be in the process of becoming unbound, is minimized.   

\item We have investigated whether models which are biased (for example, due to the underlying asphericity of the system) manifest themselves through poorer fits to the data. We found that these models typically have a higher total value of the $\chi^2$ (see Fig.~\ref{fig10}). We suggest a $\chi^2_{\rm tot} = 3.5$ cut to weed out biased models and reduce the uncertainty in the recovered profiles.
\end{itemize}

In conclusion, {\sc GravSphere} is certainly a promising method for modelling dark matter distributions in dwarf galaxies. It remains to be seen whether it can maintain its lack of bias in the innermost regions of dwarfs (below $R_e/2$), which would be possible with higher-resolution simulations. In this work we focused on studying the effects of the violation of the assumptions of the spherical Jeans equation on {\sc GravSphere}'s performance. This study could, in the future, be extended with an inclusion of the effects of stellar binaries and contamination from the Galactic halo to provide a more realistic description of the expected systematics. This, together with modelling of multiple tracer populations and proper motions, has been explored on more idealized systems in \citet{gravsphere}. We aim to extend this study with $N$-body and hydrodynamics simulations of realistic dwarf spheroidals in future work.

\section*{Data availability}
The data analysed in this article can be made available upon reasonable 
request to the corresponding author. {\sc pyGravSphere} is open source software and is available on \url{https://github.com/AnnaGenina/pyGravSphere}.

\section*{Acknowledgements}
We thank the anonymous referee for suggestions that significantly improved this paper. We are also grateful to Sebastiaan L. Zoutendijk for his contributions to the {\sc pyGravSphere} code. This work was supported by the Science and Technology Facilities Council (STFC) consolidated grant ST/P000541/1. AG acknowledges an STFC studentship grant
ST/N50404X/1. CSF and SC acknowledges support by the European Research
Council (ERC) through Advanced Investigator grant DMIDAS (GA 786910). This work used the DiRAC Data Centric system at Durham University,
operated by the Institute for Compu
tational Cosmology on behalf of the
STFC DiRAC HPC Facility (\url{www.dirac.ac.uk}). This equipment was
funded by BIS National E-infrastructure capital grant ST/K00042X/1,
STFC capital grant ST/H008519/1, and STFC DiRAC Operations grant
ST/K003267/1 and Durham University. DiRAC is part of the National
E-Infrastructure. ADL is supported by the Australian Research Council (project Nr. FT160100250). AR is supported by the European Research Council's Horizon2020 project `EWC' (award AMD-776247-6). This work has benefited from the use on {\sc numpy}, {\sc scipy} and {\sc matplotlib}.




\bibliographystyle{mnras}
\bibliography{beta.bib} 




\appendix

\section{Convergence criteria and generating initial positions}
\label{convergence}

\begin{figure}
\centering
		\includegraphics[width=0.9\columnwidth]{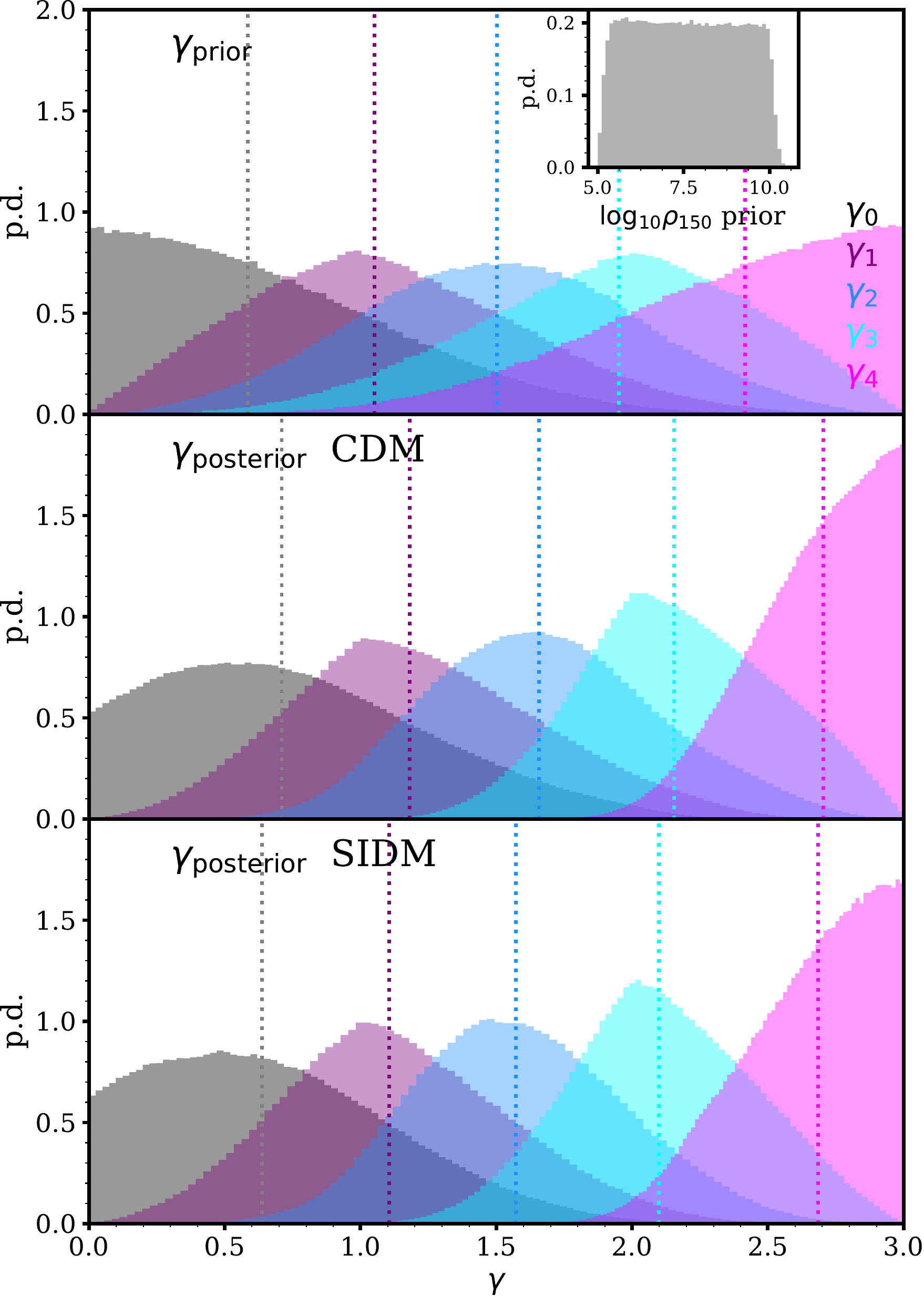}
		\caption{\textit{Top:} effective priors on the logarithmic slopes $\gamma_j$. Each slope is identified by its colour. The inset shows priors on the density at 150~pc, assuming a 1~kpc half-light radius. \textit{Middle:} posterior distributions, collected from all CDM dwarfs within our sample, weighted equally. The dotted lines are the medians of the distributions. \textit{Bottom:} posterior distributions, collected from all SIDM dwarfs within our sample, weighted equally. The dotted lines are the medians of the distributions.}
		\label{AppFig1}
\end{figure}
	\begin{figure}

		\includegraphics[width=\columnwidth]{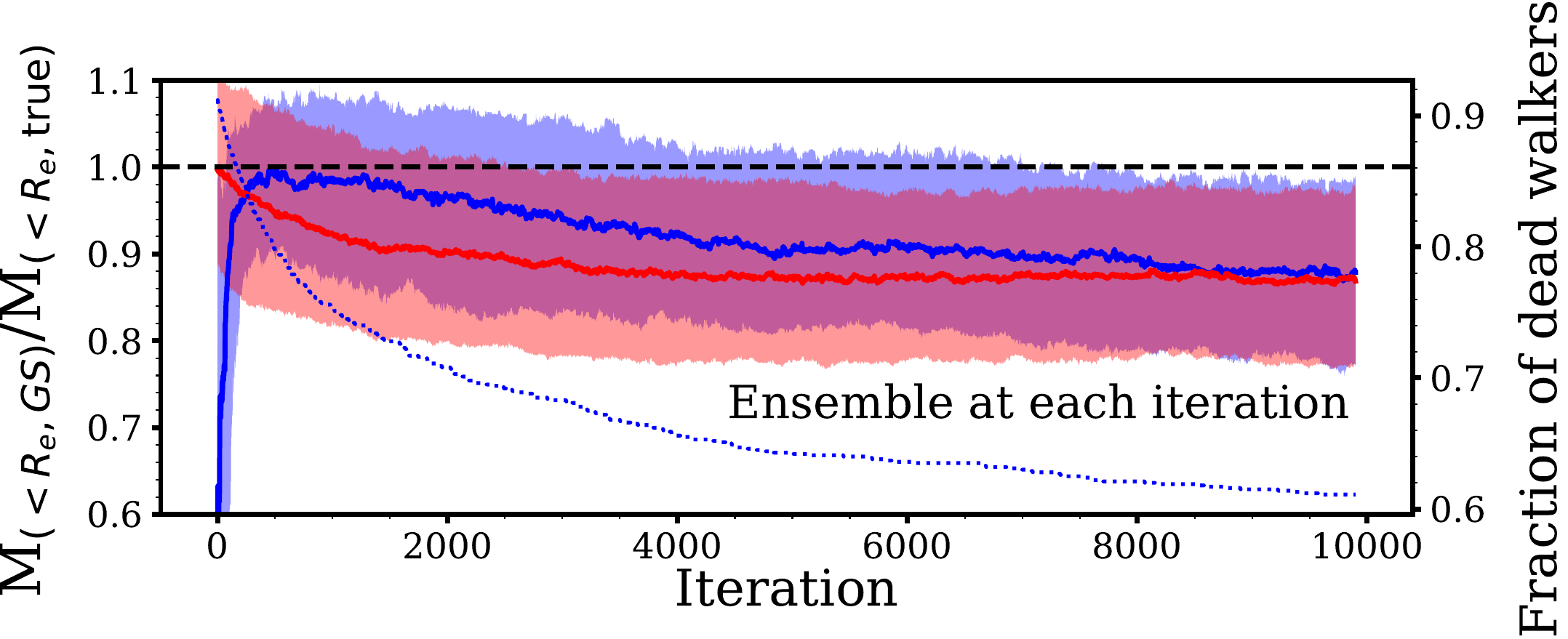}

		\caption{The convergence of mass within the half-light radius, expressed as a ratio of this mass to the true value, for two different prior selection methods. Priors that are generated completely uniformly are shown in solid blue, and those selected in a uniform, yet conditional, fashion (such that all priors satisfy the monotonic increase in the values of $\gamma_j$) are shown in red. The shaded bands represent 68 per cent confidence limits. The blue dotted line shows the fraction of `dead walkers' (those stuck in the infinitely negative log-likelihood space), when using fully uniform priors. }

		\label{AppFig2}
\end{figure}
	\begin{figure*}
        \centering
		\includegraphics[width=2\columnwidth]{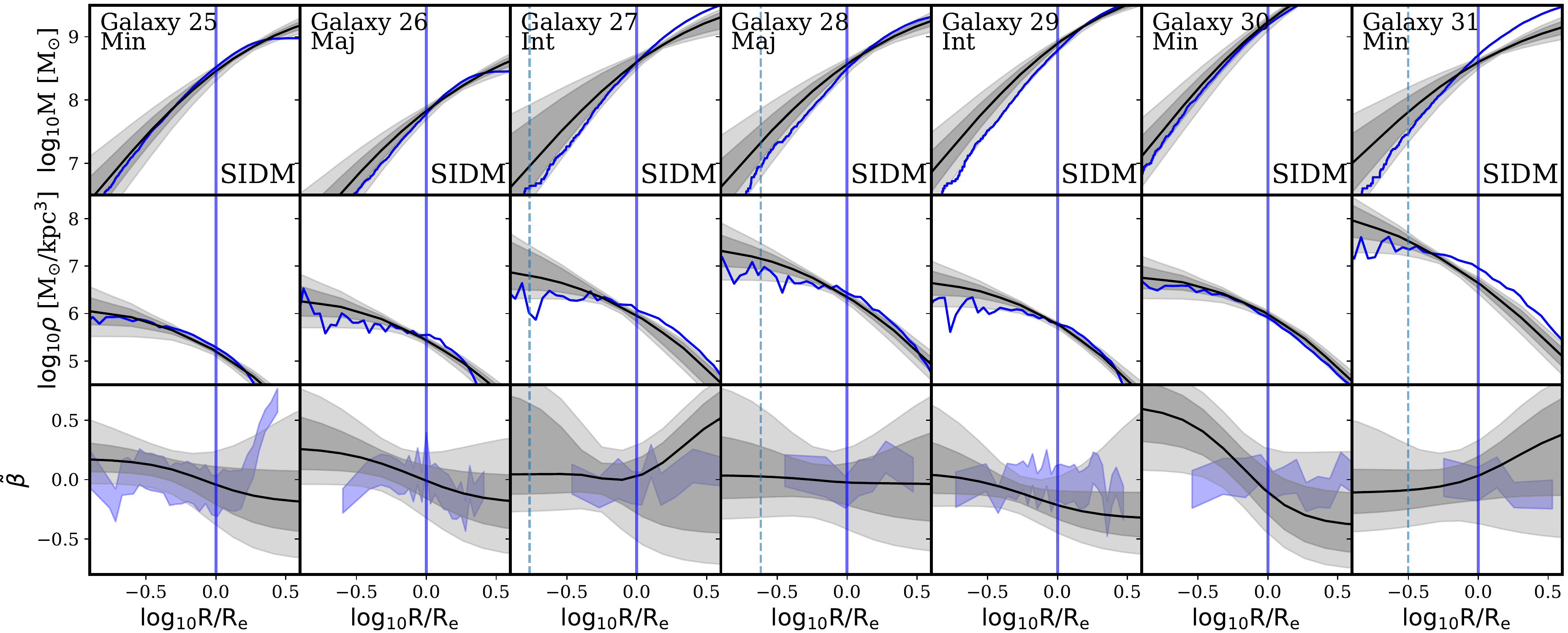}

		\caption{As Fig.~2c, but now using an extended set of priors on the density slopes $\gamma_j$, allowing `holes' in central regions of dwarfs. Note that for a number of these dwarfs the spatial resolution, $2.8\epsilon$ (vertical dashed line), is below 0.125$R_e$ and below the limits of the figure. }

		\label{AppFig3}
\end{figure*}

In this section we describe the effect of our choice of the initial positions of {\sc emcee} walkers on the convergence of {\sc GravSphere}'s results. 

\subsection{Effective priors}

As mentioned in the main text, we generate the initial positions through the selection of walkers that satisfy the condition of radial increase in power-law slopes, $\gamma_j$, and the constraint on smoothness, $\Delta\gamma=1$.

The top panel of Fig.~\ref{AppFig1} shows the effective priors on each slope. It is clear that these priors are not uniform as the selection of the slopes is not independent; however, the width of these distributions allows for a variety of density profiles. The inset shows the priors for the density at 150~pc, $\rho_{150}$. Despite the non-uniform nature of the $\gamma_0$ prior, when combined with a uniform prior on the scale density, $\rho_0$, the resulting $\rho_{150}$ prior is effectively uniform and is not biased towards more core or cusp-like values \citep{heating}.

The middle panel of Fig.~\ref{AppFig1} suggests that the posteriors on $\gamma$ are not completely determined by the priors. The prior and posterior distributions are offset, as seen from their median values, and their shapes are noticeably different. The $\gamma_4$ posterior is clearly pushing against the prior boundary, suggesting that a wider prior on this parameter is desirable. 

In the bottom panel, we see similar posterior distributions for SIDM dwarfs. It is clear that the regions with $\gamma_0 = 0$ are not prioritized by the {\sc emcee} walkers, resulting in an inference of more cuspy dark matter density profiles. It can be seen, however, that lower values of $\gamma_j$ are preferred, compared to the CDM sample.

\subsection{Walker convergence compared to previous implementations}

Previous implementations of {\sc GravSphere} have used initial positions of the walkers for the broken power law slopes $\gamma_j$ that are completely uniform. This results in a large number of {\sc emcee} walkers starting off in regions of infinitely negative log-likelihood. This is because these walkers do not satisfy the constraints for monotonically increasing values of $\gamma_j$. We will refer to these as `dead walkers'. Eventually, some of these climb out and explore the posterior distribution, but not all, and this can take many iterations. In \citet{gravsphere}, the chains were run for 5000 iterations, with the last 2500 used for analysis. In Fig.~\ref{AppFig2} we compare this method to one employed in this work. 

We pick Galaxy~4 as our representative example and we select the mass within the half-light radius, $M(<R_e)$, as a quantity for which we wish to establish convergence. Fig.~\ref{AppFig2} shows the median value of the bias, $M(< R_e)$/$M_{\rm true}$, and the 68 per cent confidence levels for \textit{each} walker iteration using the original {\sc GravSphere}'s method for generating initial positions (blue) and the initial positions generated using the method described in this work (red). The blue dotted line shows the fraction of dead walkers remaining after each iteration when using the original {\sc GravSphere} method (right vertical axis).

 It can be seen that our new method reaches convergence after $\sim5\times10^{3}$ iterations, whereas the original method requires $\sim$500 iterations to get out of the low log-likelihood regions and, in fact, does not reach the converged distribution until after $\sim9\times10^{3}$ MCMC iterations. The chains start off with over 90 per cent dead walkers. This percentage drops to $\sim 60$ per cent near 10$^{4}$ iterations and can be seen to decrease slowly. We conclude that our new method of initial position selection allows for faster chain convergence and an efficient walker exploitation.

\subsection{Extended priors that favour cores}

In Fig.~\ref{AppFig3} we display the mass, density and anisotropy profile recovery with {\sc GravSphere} for our sample of SIDM dwarfs when the priors on the power law density slopes $\gamma_j$ are allowed to vary between $\gamma_j=[-2,3]$, allowing `holes' in the density distribution. In the post-processing, we fix all slopes $\gamma_j< 0 $ to $\gamma_j = 0$. This effectively increases the sampling by {\sc emcee} walkers of the parameter space regions where the dark matter distribution is cored.

\bsp	
\label{lastpage}
\end{document}